\documentclass[preprint]{aastex}

\def\simlt{\lower.5ex\hbox{$\; \buildrel < \over \sim \;$}}
\def\simgt{\lower.5ex\hbox{$\; \buildrel > \over \sim \;$}}

\def\cm3{{\rm cm^{-3}}}
\def\kms{km s$^{-1}$}
\def\msol{{M$_\odot$}}

\def\vlsr{{v_{\rm LSR}}}

\def\schi{{\sc Hi}\ }
\def\dkpc{{d_{\rm kpc}}}
\def\degs{{$^\circ$}}
\lefthead{Koo et al.}
\righthead{ HI STUDY OF SOUTHERN GALACTIC SUPERNOVA REMNANTS}

\begin{document}

\title{ HI 21 CM EMISSION LINE STUDY OF SOUTHERN GALACTIC SUPERNOVA REMNANTS}

\author{Bon-Chul Koo and Ji-hyun Kang}
\affil{Astronomy Program, SEES, Seoul National University,
Seoul 151-742, Korea;\\
koo@astrohi.snu.ac.kr}

\author{N. M. McClure-Griffiths}
\affil{ATNF-CSIRO, PO Box 76, Epping NSW 1710, Australia;\\ Naomi.McClure-Griffiths@atnf.csiro.au}

\begin{abstract}

We have searched for 
\schi 21~cm line emission from shocked atomic gas associated with 
southern supernova remnants (SNRs) using 
data from the Southern Galactic Plane Survey.
Among the 97 sources studied, we have detected 10~SNRs with high-velocity 
\schi emission confined to the SNR.
The large velocity and the spatial confinement suggest that 
the emission is likely from the gas accelerated by the SN blast wave. 
We also detected 22~SNRs which show \schi 
emission significantly brighter than the
surrounding regions over a wide ($>10$~\kms) velocity interval. 
The association with these SNRs is less certain. 
We present the parameters and maps of the excess emission in these SNRs. 
We discuss in some detail the 
ten individual SNRs with associated high-velocity \schi emission.

\end{abstract}

\section{INTRODUCTION}

The {\sc Hi} emission line at 21~cm is very useful for studying 
radiative shocks in old supernova remnants (SNRs). The line is emitted
from cold atomic gas, which is formed when the shock becomes radiative.
This occurs when the shock has swept up enough ($\simgt  10^{19}$~cm$^{-2}$)
column density. By analyzing {\sc Hi} emission lines from old SNRs,
we can derive some fundamental SNR parameters, such as 
the shock velocity, mass and kinetic energy of the
shocked gas, age of the SNR, etc. 
In addition, the {\sc Hi} data reveal the  distribution of
shocked atomic gas, which can be compared with the distribution of
ambient atomic and molecular gas to explore the details of the interaction
of SNRs with the ambient interstellar medium (ISM). 
Understanding such details, and in particular
for a reasonably large statistical sample, is crucial for understanding
the supernova process itself and also 
the overall effects of supernovae on the ISM.
    
Observationally, however, it is difficult to detect the {\sc Hi} 
emission line  from shocked atomic gas, because most known SNRs are
located in the Galactic plane where the Galactic background {\sc Hi} emission
causes severe contamination. Only when the shock velocity is 
very large (e.g., $\simgt 100$~\kms) is the emission 
clearly discernible from the background emission.
In the 1980's, high-velocity (HV) emission 
was detected toward SNRs 
IC 443, G78.2+2.1, CTB 80, CTB~109, and VRO~42.05.01
(DeNoyer 1978; Giovanelli \& Haynes 1979; 
Landecker, Roger, and Higgs 1980; Braun \& Strom 1986; Koo et al. 1990).
A systematic search of SNRs was made by Koo \& Heiles (1991; hereafter KH91)
using the Hat Creek 25~m
telescope (FWHM$=36'$). 
They observed 103 northern SNRs 
using the Hat Creek 25~m telescope (FWHM$=36'$),
which were all of the known SNRs at 
$\delta\simgt -38.5^\circ$ at that time. 
%\footnote{ After the KH91 survey in 1991, more SNRs have been
%identified in radio continuum and in X-ray, and the number 
%increased from 103 to 152 according to the Green's catalog [Green 2003].}
They observed each SNR at 9 points in a cross
pattern centered at its catalog position and searched for broad excess 
emission brighter than the surrounding region.  
The SNRs with excess emission are divided 
into 3 ranks, in which increasing number implies
increasing reliability of the detected \schi feature. 
Fourteen SNRs 
(excluding G117.4+5.0 which is no longer considered an 
SNR any more [Green 2001])
showed HV excess emission which is quite
likely from the gas accelerated by the SNR shock and ranked 3. 
The classical source IC~443 was classified as rank 2 because of some 
confusing emission in the surrounding region. 
%\footnote{These 15 SNRs include
%the fourteen rank 3 SNRs (excluding G117.4+5.0 which is not considered as a 
%SNR any more [Green 2001]) and the classical source IC~443.} 
The velocity of \schi gas ranged from 70 to 160~\kms\ with
respect to the systemic velocities.
Follow-up high resolution observations have been made for 
SNRs CTB 80, W44, and W51C, and 
confirmed the association
(Koo et al. 1993; Koo \& Heiles 1995; Koo \& Moon 1997). 
 
	In this paper, we present the results of a systematic search 
for shock-accelerated HV \schi gas toward 97 SNRs 
in the southern sky. 
The organization of the paper is as follows: In \S~2, we 
describe the data and the procedure for classifying SNRs. 
We divide the SNRs into three ranks in a way similar to KH91. 
In \S~3, we present the tables and maps summarizing the results. 
We have detected ten SNRs with HV \schi gas
localized at the positions of SNRs and 
another 22 SNRs with some excess emission at the position of SNRs. 
In \S~4, we discuss ten individual SNRs, and,
in \S~5, we summarize the paper.

\section{DATA AND THE CLASSIFICATION}

	We use Parkes data from the Southern Galactic Plane Survey 
(SGPS; McClure-Griffiths et al. 2001) for the
search. The data covers 
$l=253^\circ$--$358^\circ$ and $|b|\simlt 10^\circ$ with an angular 
resolution of 16$'$ (pixel size 4$'$) and a velocity resolution of 0.82~\kms.
The sensitivity ranges from 0.13 to 0.27 K, but it is much better 
(0.06--0.09 K) in the latitude range
$-1.^\circ 5 \simlt b \simlt 1.^\circ 5$.
With a sensitivity of
$\sim$0.1~K (1.5$\sigma$), the minimum detectable \schi mass within a  
10~\kms\ velocity interval is $\sim 0.4 \dkpc^2$~\msol, 
where $\dkpc$ is the distance to the source in kpc. 

There are 97 Galactic SNRs 
in Green's catalog (Green 2001) that lie 
in the longitude and latitude range covered by the SGPS data.  
We compare the average \schi spectra of these SNRs to those of 
the surrounding regions, and 
look for excessive emission wider than 10~\kms\ and 
localized at the position of SNRs. 
Two circular/elliptical shells surrounding the SNR 
are used for the "OFF" spectra:
one which is $8'$ (half-beam width) thick and 
has an inner boundary at $8'$ apart from 
the SNR boundary, and the other 
which is also $8'$ thick and $8'$ from the outer boundary of the 
former shell.
We divide the SNRs into 3 ranks,
where the increasing number implies increasing reliability of the detected H~I
feature:

\noindent
Rank 2: SNRs which have an excessive emission over the outer shell by 
more than the $1.5$ times of its rms fluctuation
over a wide ($\ge 10$~\kms) velocity interval. \hfill\break
Rank 3: SNRs which have an excessive emission over the inner and outer shells 
at extreme velocities. \hfill\break
All other sources are ranked 1.
Figs. 1 and 2 show the spectra of ranks 2 and 3 sources, respectively.

Our classification is similar to the KH91's in principle, e.g.,
our ranks 2 and 3 correspond to their ranks
2 and 3, respectively, while our rank 1 includes their ranks 0 and 1. But, since
we used a shell pattern instead of a cross pattern for the ``OFF" spectra, the 
assigned ranks for individual SNRs could be different (see next section).
SNRs ranked~3 have HV \schi gas
confined to the SNR. Such gas is rare; thus, it is probable that the
gas is really associated with the SNR. SNRs ranked~2 have excess \schi gas
confined to the position of SNRs, but not at an extreme velocity. 
Real association with the SNR is less certain, but the positional coincidence is
suggestive. 

\section {RESULTS}

The results of our survey are summarized in Table 1. 
Source parameters are from Green (2001).  The size is 
the angular size in radio continuum; a single value is quoted for nearly 
circular remnants, and the product of two values, the major and minor axes, 
is quoted for elongated remnants. The
type `S', `F', or `C' represents SNRs with a `shell', `filled-centre', or
`composite' radio structure. The fluxes at 1 GHz are also listed. 
The uncertain parameters are listed with a question mark.  
Of the 97 sources examined, 10 SNRs are ranked 3, 22 SNRs are ranked 2, and 
65 SNRs are ranked 1. 
Nine SNRs had been observed by KH91: G261.9+5.5 (2), 
G348.5+0.1 (2), G348.7+0.3 (2), 
G349.7+0.2 (2), G351.2+0.1 (2), G352.7$-$0.1 (2), G355.9$-$2.5 (0), 
G357.7$-$0.1 (2), and G357.7+0.3 (2), 
where the numbers in parenthesis are the ranks assigned by KH91. We note 
that, for G261.9+5.5, G348.5+0.1, G348.7+0.3, and G357.7$-$0.1, 
we assign a rank of 1, which KH91 assigned a rank of 2.
The disagreement is probably due to 
different observing patterns, e.g., shell versus cross. 

In Table~2 we list, for rank 2 SNRs, 
the velocity intervals at which the SNR shows excessive emission 
brighter than $1.5 \sigma$ and wider than 10 \kms\ 
and the average integrated
antenna temperature of the excess emission over the SNR area. 
The intensity of the excess emission ranges from 18 to 660~K \kms, or 
0.03 to 1.2$\times 10^{21}$~cm$^{-2}$ in \schi column density (assuming that 
the emission is optically thin).
Fig. 3 shows the maps of the excess emission where the 
dotted circles mark the positions and geometrical mean 
sizes of the SNRs. 
All rank 2 SNRs show at least some 
excess emission and 
we give a very short description of its morphology in Table~2.
SNRs which show distinct excess emission confined to a small area 
within or near their boundaries are 
G286.5$-$1.2, G296.1$-$0.5, G301.4$-$1.0, and G302.3+0.7.

Table~3 summarizes the results of rank 3 SNRs. 
The velocity intervals are approximate intervals at which the SNRs show
excessive emission compared to their surroundings.
All sources, but two 
(G320.6$-$1.6 and G347.3$-$0.5) show excess emission at 
negative velocities. 
The intensities of the 
excess emission are much weaker than rank 2 SNRs, e.g.,
2 to 34~K \kms\ or 0.3 to 6$\times 10^{19}$~cm$^{-2}$. This may indicate 
that rank 3 SNRs 
have swept up less mass and therefore relatively younger 
than rank 2 SNRs, although some excess emission toward rank 2 SNRs 
might be due to random fluctuations of the background emission.
Fig. 4 shows the maps of their excess emission. 
All sources show excess emission confined to them. The ones with 
particularly distinct excess emission are G272.2$-$3.2, G321.9$-$0.3, 
and G347.3$-$0.5. 
Individual sources are discussed in \S~4.

A characteristic of rank 3 SNRs is 
that most of them, 8 out of 10, show excess emission 
at negative velocities. We note that 
they are all in the fourth quadrant at $l\simgt 320^\circ$.  
According to KH91, the ones in the first quadrant, 12 out of 14, 
show excess emission at positive velocities. 
This highly contrasting tendency is consistent with what we would 
expect from the Galactic rotation: Suppose that there is an SNR 
which has a spherically symmetric,  
expanding \schi shell in the first quadrant in the inner Galaxy. 
Along the direction toward the SNR, there is a strong Galactic background
\schi emission over a wide velocity interval, from negative to positive 
velocities. Now since the LSR velocity of the SNR is positive, 
only the receding portion of the shell will be detectable.
For those in the fourth quadrant, it will be the opposite.
Therefore, the trend seems to suggest that most of the \schi features 
in Table 3 are produced by sources in the inner Galaxy, probably the SNRs.

Fig. 5 shows the longitude distribution of SNRs in different ranks.
The total number of observed SNRs is 189 including 
92 SNRs observed by KH91.
\footnote{KH91 observed 103 northern SNRs, nine of which are included 
in this study.
Excluding these nine SNRs and G112.0+1.2 and G117.4+5.0 which are no longer 
thought be SNRs, we are left with 92 SNRs. 
} 
For comparison, the total number of SNRs compiled by Green
in 2001 is 231 (Green 2001). 
The longitude distribution of SNRs is slightly asymmetric with a relatively small 
number of SNRs between $l=240$\degs\ and 290\degs. 
%This probably indicates that the identification
%of SNRs in the southern sky is relatively incomplete. 
Out of 189 SNRs, 29 SNRs are classified as rank 3, 
39 SNRs as rank 2, and the rest as rank 1. 
Fig. 5 shows that 
90\% of rank-2 SNRs are between $l=290$\degs\ and 50\degs, while it is 73\% 
for all the observed SNRs. This crowding of rank 2 SNRs in the inner Galaxy
was noted by KH91 who attributed
it to the confusion, i.e.,
the long path length gives a high probability for the random fluctuations
of the background emissions to produce rank-2 sources.
There is no such crowding of rank 3 SNRs.

\section {DISCUSSION OF INDIVIDUAL 
SUERPNOVA REMNANTS WITH HIGH-VELOCITY \schi GAS}

In the following we discuss the individual rank 3 SNRs that are 
given listed in Table 3. 
Three SNRs, 
G272.2$-$3.2, G321.9$-$0.3, and G347.3$-$0.5, are discussed in 
detail, while the other SNRs are discussed very briefly.

\subsection {G272.2$-$3.2}

G272.2$-$3.2 was discovered in the ROSAT All-Sky Survey \citep{gre93} 
and the follow-up 
X-ray and radio observations confirmed that it was a SNR \citep{gre94, dun97}.
The SNR appears centre-filled both in radio and 
in X-rays and is almost circular with a diameter of about 15$'$ 
\citep{dun97,har01}.
%There are bright radio and X-ray ``blobs" embedded within the diffuse emission. 
%They are not correlated, but the radio blobs are associated with 
%bright optical filaments \citep{win93,dun97}.
The spectral index ($0.55\pm0.15$) of the radio emission is steeper 
than plerions and the X-ray spectrum is thermal. 
In a detailed X-ray study, \cite{har01} found that the X-ray emission from 
the SNR is best described by a nonequilibrium ionization model with a 
temperature around 0.7 keV, ionization timescale of 3200~cm$^{-3}$ yr, and
absorbing column density of $1.1\times 10^{22}$~cm$^{-2}$. 
Those authors also discovered an infrared shell  
surrounding the western boundary of the SNR 
in the IRAS 60/100~$\mu$m surface brightness map. The distance to the SNR is 
uncertain. 
\cite{gre94} determined a distance of $1.8^{+1.4}_{-0.8}$~kpc by comparing their  
X-ray absorbing columns ($4.6\pm 3 \times 10^{21}$~cm$^{-2}$) 
to a mean interstellar absorbing columns ($2.5\times 10^{21}$~cm$^{-2}$ kpc$^{-1}$).
As they pointed out, this distance would imply an initial SN explosion energy of 
$2.5\times 10^{49}$~erg based on the Sedov model, which is uncomfortably small.
On the other hand, \cite{har01} derived an upper limit of 10 kpc by 
comparing their X-ray absorbing columns ($1.12\pm 0.02 \times 10^{22}$~cm$^{-2}$)
to the interstellar absorbing columns ($1.2\times 10^{21}$~cm$^{-2}$ kpc$^{-1}$) 
derived by \cite{luc78} toward this direction, and 
adopted a distance of 5 kpc. 

The excess \schi emission toward G272.2$-$3.2 
appears as a faint ($\simlt 0.3$~K) bump 
at $\vlsr=-94$ and $-70$~\kms (Fig. 2).
Fig. 6 shows the contour map of the integrated intensity 
overlaid on the X-ray image 
of G272.2$-$3.2. The \schi emission peaks at the
western portion of the remnant where the infrared shell has been detected. 
The X-rays are enhanced at the western edge of the SNR 
and the SNR boundary is somewhat flattened at this location.
All these observations seem to indicate that 
G272.2$-$3.2 has been interacting with a dense medium in the west,  
but
the almost circular shape of the SNR suggests that the interaction 
might have started rather recently.
Also, the detection of the HV \schi gas 
suggests that the SNR shock in this region is radiative.
At a distance of 5 kpc, the radius ($7.5'$) 
of the SNR is $11$ pc and
its systematic velocity is 
27~\kms\ 
(assuming a flat rotation curve with $R_\odot=8.5$~kpc and $v_\odot=220$~\kms). 
The maximum velocity of the \schi gas that we detected corresponds to 
an expansion velocity of $v_s\simeq 120$~\kms\ with 
respect to the SNR. 
%On the other hand, the temperature of the X-ray emitting hot 
%gas is $0.73$~keV \citep{har01}. 
%The centrally-peaked thermal X-ray emission suggests that the hot gas might be 
%roughly isochoric and the evaporation model of \cite{whi91} may be applicable.
%In this model, the observed X-ray temperature is $\simeq 0.72 T_s$ where $T_s$ is 
%the postshock temperature. (See Koo, Kim, \& Seward 1995.) 
%Therefore, the implied shock velocity 
%is 660~\kms. (We assume a fully-ionized preshock gas.) 
%If we assume that the driving pressures for the X-ray and atomic shocks are 
%comparable, then the preshock density of the latter is  
%$\sim (660/120)^2=30$ times higher than that of the former. 
%\cite{har01} derived a mean density of 0.23--0.28~cm$^{-3}$ (adopting 15$'$ instead of $16'$), 
%which suggests that the preshock
%density for the atomic shock is 7--8~cm$^{-3}$.

\subsection {G321.9-0.3}

G321.9$-$0.3 is a shell-type SNR with a  bright western shell. 
\cite{kes87} classified the SNR 
as one of the barrel-shaped SNRs, which are the SNRs 
with two symmetric bright limbs with little or no emissivity in the end-caps. 
No reliable distance estimates are available. Barring any other distance 
indicators, one can use the empirical 
$\Sigma$-$D$ relation to estimate the distance to within about a factor of two.
Using the relationship given by \cite{cas98}, 
we estimate a distance of 5.5 kpc (cf. Clark, Caswell, \& Green 1975), 
but note that this is 
very uncertain. At this distance, 
the major and minor axis of the SNR become 50 pc and 37 pc, respectively. 
G321.9$-$0.3 is particularly interesting because 
there is a X-ray binary, Circinus X-1, located $10'$ north of 
the northern rim of the SNR. It has been proposed that Circinus X-1 is 
a runaway binary ejected from the SN explosion \citep{cla75b, ste93}.
The minimum distance to Cir X-1 from \schi absorption experiment is 
6.7 kpc \citep{gos77}.

G321.9$-$0.3 is a prototype of rank 3 sources.
It has a 
broad ($\sim 50$~\kms) wing clearly confined to the SNR.
Fig. 7 shows the contour map of the integrated intensity 
between $\vlsr=-162$ and $-109$~\kms\ 
overlaid on the radio
continuum image of the SNR. The \schi emission peaks at the 
very center of the remnant. 
If the distance to the SNR is 5.5 kpc, its 
systemic velocity would be $-78$~\kms\ and 
the maximum velocity of the \schi gas corresponds to 
an expansion velocity of $\sim 80$~\kms. 
The characteristic age of the remnant, assuming that it is in a 
pressure-driven snowplow phase, is 
$\sim 8\times 10^4$~yr. 
If Cir X-1 was ejected from the center of G321.9$-$0.3, then, the current distance 
($25'=40$~pc) implies a transverse speed of 490~\kms.

\subsection {G347.3$-$0.5}

G347.3$-$0.5 (RX J1713.7$-$3946) is 
bright X-ray SNR discovered in the ROSAT 
All-Sky Survey (Pfeffermann \& Aschenbach 1996). It has a bright  
western shell and an unidentified point source at the center.
The X-ray spectrum of the diffuse emission is 
non-thermal, which indicates the presence of high-energy electrons 
accelerated by the SNR shock \citep{koy97, sla99, uch03, laz04}.
TeV gamma-ray emission has been detected toward the X-ray bright 
western shell, which is attributed to 
the inverse Compton scattering of the Cosmic Microwave Background 
Radiation by high-energy electrons (Muraishi et al. 2000) or 
pion decay (Enomoto et al. 2002).
At radio wavelengths, the SNR is generally 
very faint with a few relatively bright arc-like features 
along the west rim (Slane et al. 1999).
The distance to the SNR is controversial: Koyama et al. (1997) adopted 1 kpc 
by comparing the X-ray derived absorbing columns ($6\times 10^{21}$~cm$^{-2}$)
to the average total ($6\times 10^{22}$~cm$^{-2}$) columns in the direction of 
the Galactic center. This gives a radius of 9 pc and suggests that 
the SNR is young, 
possibly the remnant of the historical SN AD 393 (Wang et al. 1997). 
But Slane et al. (1999) pointed out that the total column density 
toward the SNR direction ($1.2\times 10^{22}$~cm$^{-2}$) is significantly lower
than the average in this direction, so that the distance to the remnant must be 
considerably larger than 1 kpc. They further argued that the SNR is possibly 
associated with molecular clouds in the vicinity, which have central velocities 
of $-94$~\kms\ and/or $-69$~\kms, and adopted 6 kpc. This gives the SNR radius of 
60 pc and suggests that the SNR old. \cite{koo04} showed how the column 
density of hydrogen nuclei (\schi+H$_2$) varies with distance toward the 
SNR using a kinematic model of the Galaxy 
and showed that the accumulated column density becomes comparable to 
the X-ray absorbing columns at $\sim 1 $~kpc.

The excess \schi emission toward G347.3$-$0.5 is detected between
$\vlsr=68$ and $81$~\kms\ at the western rim where there is 
enhanced X-ray emission.
Fig. 8 shows ROSAT X-ray image of the SNR with the total integrated \schi 
intensity contours overlaid.
At a distance of 1 kpc, the radius ($30'$) of the SNR is 9 pc 
and the systematic velocity of the SNR would be $\sim -6$~\kms.
The maximum velocity of the \schi gas corresponds to 
an expansion velocity of $\sim 90$~\kms. 
The X-ray observations, however, showed that 
thermal X-rays from shocked gas are absent or at least very weak, 
which has been considered to be due to very tenuous ambient medium
\citep{sla99}. The SNR, then, must be very young and the shock should be 
non-radiative. The HV \schi gas could still have been produced by 
SNR shocks propagating through ambient molecular clouds.
Indeed there are molecular clouds surrounding the 
SNR at $\vlsr\sim -10$~\kms, which have been suggested to be 
interacting with the SNR \citep{fuk03}. 
Alternatively, if 
the SNR is at a distance of 6 kpc and interacting with a molecular cloud at 
$\vlsr=-69$~\kms\ along the western rim as \cite{sla99} suggested, 
then the fast-moving \schi gas could have been produced by the interaction
with this cloud. 

\subsection{Other Rank-3 SNRs}

G320.6$-$1.6 is a shell-type SNR composed of multiple arcs 
\citep{whi96}. Only the southern and eastern parts are visible, and 
the extent of the remnant appears to be less than a degree.
Excess \schi emission,
which attains maxima near the northern and southern 
boundaries of the remnant, is apparent
at positive LSR velocities, $\vlsr=101$ to 121~\kms.
There is some confusing emission in the surrounding area, and 
the association is uncertain.

G332.4+0.1 (or MSH 16--51, Kes 32) is a shell-type SNR with peculiar 
morphology \citep{whi96}. The overall shape is circular with diameter of 
15$'$, but the eastern shell protrudes beyond the circular 
boundary. From the north of the protruding structure
a jet-like feature is emerging and expanding into a 
plume of low-brightness emission \citep{rog85, whi96}. 
There is a radio source at the center, but no pulsation has been discovered. 
X-rays have been detected toward the SNR \citep{bri99}, 
but the association is not clear. 
There is an excess \schi emission which attains 
a maximum toward the SNR. But it appears to be connected to 
a stronger, more extended excess emission 
to the northwest beyond the SNR, and
the association with the SNR is uncertain.  

G335.2+0.1 is a shell-type SNR of $21'$ diameter 
with enhanced emission at southwest \citep{whi96}. A pulsar, offset from the remnant 
center by $\sim 4'$, has been discovered \citep{kas96}. Its large 
characteristic age ($2.7\times 10^6$~yr), however, suggests that the 
association with the SNR is unlikely, although it is not impossible if 
the pulsar was born with a large spin period.
The excess \schi emission peaks in the southwest portion of the SNR
where the radio continuum is enhanced. The morphological correlation 
is suggestive of their association.

G338.5+0.1 is a $9'$-size SNR with amorphous morphology 
\citep{whi96}. The extent to the south is unclear because it  
partially overlaps with a bright HII region.
Very weak excess \schi emission is discovered around the SNR.
The emission, however, 
is extended over an area larger than the SNR, and 
the association is not clear.

G343.1$-$2.3 was originally discovered by \cite{mca93} as an SNR associated 
with the young ($\sim 1.7\times 10^4$~yr) 
gamma-ray pulsar PSR B1706$-$44 (PSR J1709$-$4428). But the 
association was questioned later based on several arguments including 
distance inconsistencies 
(Frail, Goss, \& Whiteoak 1994; Nicastro, Johnston, \& Koribalski 1996; 
see also Giacani et al. 2001).
The SNR appears as a faint, incomplete shell with a diameter of $\sim 40'$.
There is strong ($\sim 2$~K), 
excess \schi emission which attains a maximum within the western boundary of 
the remnant, but there is some confusing emission in the surrounding area.

G343.1$-$0.7 is a square-shaped, shell-type SNR with bright, 
$\sim 20'$-long southern and eastern sides \citep{whi96}.
The northern and western parts are faint and 
overlap with a smaller ($\sim 10'$) shell which is probably thermal \citep{whi96}.
Excess \schi emission is discovered within the western boundary of 
the remnant. It is superposed on a larger structure crossing western 
portion of the remnant, and the association is uncertain.

G350.0$-$2.0 is a shell-type SNR consisting of filaments with 
very different curvatures and brightnesses \citep{gae98}. To the northwest 
it has a bright, $40'$-long filament of small curvature, while, to southeast,  
it has a faint circular filament of $\sim 30'$ diameter. 
The whole extent of the SNR is $\sim 45'$. 
\cite{gae98} classified it as ``bilateral" SNR, and 
considered several mechanisms for the noncircular morphology,  
including multiple SN explosions and an explosion in a stratified ISM.
We have detected very weak excess \schi emission
in the northwestern portion of the remnant.
This is where the SNR shell is bright, and it is possible that the SNR is 
interacting with a dense material in this direction.
The SNR shock propagating into a dense medium would be radiative, and the 
association of the excess \schi emission with the SNR is quite possible.

\section {SUMMARY}

We have studied \schi emission lines toward 97 southern SNRs 
between $l=253^\circ$--$358^\circ$ and $|b|\simlt 10^\circ$ 
using the SGPS data.
We compare the average \schi spectra of these SNRs to those of the surrounding regions, and 
look for excessive emission wider than 10~\kms\ and 
localized at the position of the SNRs.
We divide the SNRs into 3 ranks, where a higher
rank means higher probability of an associated \schi gas;
10~SNRs are ranked 3, 22~SNRs are ranked 2, and 65~SNRs are ranked 1.
SNRs ranked~3 have HV \schi gas
confined to the SNR. Such gas is rare; thus, it is probable that the
gas is really associated with the SNR. SNRs ranked~2 have excess H~I gas
confined to the position of SNRs, but not at an extreme velocity. 
Real association with the SNR is less certain.
All other sources are ranked 1.

We present the parameters and maps of the excess emission associated with rank 2 and 3 SNRs. 
All rank 2 SNRs show at least some 
excess emission, but the following ones show 
particularly distinct enhancements:  
G286.5$-$1.2, G296.1$-$0.5, G301.4$-$1.0, and 
G302.3+0.7.
All ranks 3 sources show excess emission confined to them. The ones with 
particularly distinct excess emission are: G272.2$-$3.2, G321.9$-$0.3, 
and G347.3$-$0.5. These three SNRs are discussed relatively in detail. 
The other rank 3 SNRs are briefly discussed. 

High-resolution observational studies are needed 
for understanding the nature of the excess \schi emission detected in this paper.
In particular,
recent molecular line studies showed that HV \schi gas in 
some SNRs are produced by the interaction with molecular clouds, e.g., 
W44, W51C, and IC 443 (see Koo 2003 and references therein).
Therefore, both \schi and molecular line observations are essential.

\acknowledgements
The ROSAT X-ray images in Figs. 6 and 8 were retrieved from {\it SkyView} and 
we acknowledge the use of NASA's SkyView facility
(http://skyview.gsfc.nasa.gov) located at NASA Goddard Space Flight Center.
This work was supported by S.N.U. Research Fund (3345-20022017).

{}
\clearpage

\figcaption[]{
\schi spectra of rank 2 sources. The thick solid line is the 
average spectrum toward the SNR, the thin solid line is the
difference spectrum of the SNR using the outer shell 
for the ``OFF" spectrum, and the dotted line 
is the amplitude ($1.5 \sigma$) of the background fluctuation
obtained from the OFF region.
}

\figcaption[]{
\schi spectra of rank 3 sources. The thick solid line is the 
average spectrum toward the SNR, while the thin solid and dotted 
lines are the difference spectra of the SNR using the inner and 
outer shells for the ``OFF" spectrum, respectively.
}

\figcaption[]{
Contour maps of the excess emission of the rank-2 SNRs.
The contours are labeled by average antenna temperature in K.
The grey scale plot distinguishes between
valley and hill, and the higher integrated intensity is blacker. The dashed circles
represent the locations and sizes of SNRs. The source
name and the velocity range appears at the top of each picture.}

\figcaption[]{
Same as Fig. 3, but for rank-3 SNRs.}

\figcaption[]{
Distribution of SNRs in Galactic longitude. 
The histograms in solid line show the distribution of ranks 3, 2, and 1
SNRs from bottom to top in cumulative way, e.g., the sum 
under the top histogram is the total number (189) of observed SNRs. 
The top histogram in dotted line shows the distribution of 
all 231 SNRs in Green's catalog (Green 2001).}

\figcaption[]{
Contour map of the \schi excess emission
overlaid on the ROSAT X-ray image of G272.2$-$3.2. 
The contour levels are 0.1, 0.15, and 0.2~K in average antenna 
temperature over $\vlsr=-94$ to $-70$~\kms.
}

\figcaption[]{
Contour map of the excess \schi emission overlaid on the 843~MHz radio
continuum image of  G321.9$-$0.3. The radio image is from 
the MOST SNR catalog \citep{whi96}. 
The intensity has been integrated over $\vlsr=-162$ to $-109$~\kms.
The contour level starts at 0.1~K 
and increases by 0.05~K in average antenna 
temperature over $\vlsr=-162$ to $-109$~\kms.
}

\figcaption[]{
Contour map  of the excess \schi emission overlaid on the 
ROSAT X-ray image of G347.3$-$0.5. The contour level starts at 0.4~K 
and increases by 0.2~K in average antenna 
temperature over $\vlsr=68$ to $81$~\kms.
}

\clearpage

\begin{deluxetable}{clcccccc}
\tabletypesize{\scriptsize}
\tablecaption{Summary of Survey Results \label{tbl-1}}
\tablewidth{0pt}
\tablehead{
\colhead{}	&\colhead{} 	&\colhead{}	&\colhead{}	&
\colhead{}	&\colhead{}	&
\colhead{Flux at} &\colhead{}\\
\colhead{Galactic} &\colhead{} &\colhead{$\alpha$(2000)} &\colhead{$\delta$(2000)} &
\colhead{Size}	&
\colhead{} 	&\colhead{1 GHz}	&\colhead{}  \\
\colhead{Coordinates} & \colhead{Name(s)} &
\colhead{(h m s)}&\colhead{ \ ($\arcdeg$ \ $^\prime$)}&
\colhead{(arcmin)} & \colhead{Type}
& \colhead{(Jy)} & \colhead{Rank}\\
\vspace{-0.3cm}
}
\startdata

%\newcolumntype{.}{D{.}{.}{-1}}
%\begin{table}[!p]
%\scriptsize
%\begin{center}
%\caption{Summary of Survey Results }
%\vspace{0.3cm}
%\begin{tabular}{clcccc.c}
%\hline\hline
%
%& &&&&&$\rm{Flux at}$ &\\
%Galactic & hh mm nn&\arcdeg $^\prime$ & & Size & & 1 GHz &  \\
%Coordinates & Name(s) & RA(1950) & DEC(1950) & (arcmin) & Type & (Jy) & Rank
%\\
%\hline
%
260.4$-$3.4 & Puppis A, MSH 08$-$44 & 08 22 10&$-$43 00& 60$\times$50 & S & 130 \ \ \ & 1  \\
261.9+5.5 & \nodata & 09 04 20 & $-$38 42 & 40$\times$30 & S & 10? & 1 \\
263.9$-$3.3 & Vela(XYZ) & 08 34 00 & $-$45 50 & 255 \ \ & C &1750 \ \ \ \ & 1 \\
%263.9$-$3.3 & Vela(XYZ) & 08 34 00 & $-$45 50 & 255\phn\phn & C &1750 & 1 \\
266.2$-$1.2 & \nodata & 08 52 00 & $-$46 20 & 120 \ \ & S & 50? &1\\
%\vspace{0.3cm}
272.2$-$3.2 & \nodata& 09 06 50 & $-$52 02 & \ 15? & \ S? & \ \ \ 0.4 & 3  \\
%\vspace{-0.3cm}
&&&&&&&\\
279.0+1.1& \nodata &09 57 40 & $-$53 15 & 95 & S & 30? & 1\\
284.3$-$1.8&MSH 10$-$53&10 18 15 &$-$59 00 & \ 24? &S& 11?& 1\\
286.5$-$1.2&\nodata& 10 35 40 & $-$59 42 & 26$\times$6 \ & \ S? & \ \ \ \ 1.4? &2   \\
289.7$-$0.3 & \nodata& 11 01 15 & $-$60 18 & 18$\times$14 & S & \ \ \ 6.2 & 1 \\
%\vspace{0.3cm}
290.1$-$0.8 & MSH 11$-$61A & 11 03 05 & $-$60 56 & 19$\times$14 & S & 42 \ & 1  \\
%\vspace{-0.3cm}
&&&&&&&\\
291.0$-$0.1 & (MSH 11$-$62) & 11 11 54 & $-$60 38 & 15$\times$13 & C & 16 \ & 1  \\
292.0+1.8 & MSH 11$-$54 & 11 24 36 & $-$59 16 & 12$\times$8 \ & C & 15 \ & 1 \\
292.2$-$0.5 & \nodata& 11 19 20 & $-$61 28 & 20$\times$15 & S & \ \ 7? & 1 \\
293.8+0.6 & \nodata& 11 35 00 & $-$60 54 & 20 & C & \ \ 5? & 1  \\
294.1$-$0.0 & \nodata& 11 36 10 & $-$61 38 & 40 & S & $>2$? \ & 2  \\
%\vspace{-0.3cm}
&&&&&&&\\
%\hline
%\end{tabular}
%\end{center} \end{table}

%\begin{table}[!p]
%\scriptsize
%\begin{center}
%\isucontinuecaption{Summary of Survey Results }
%\vspace{0.3cm}
%\begin{tabular}{clcccc.c}
%\hline\hline
%
%& &&&&&Flux at &\\
%Galactic & & & & Size & & 1 GHz &  \\
%Coordinates & Name(s) & RA(1950) & DEC(1950) & (arcmin) & Type & (Jy) & Rank
%\\
%\hline
%
%
296.1$-$0.5 & \nodata& 11 51 10 & $-$62 34 & 37$\times$25 & S & \ \ 8?& 2  \\
296.8$-$0.3 & 1156$-$62 & 11 58 30 & $-$62 35 & 20$\times$14 & S & \ 9 & 2  \\
298.5$-$0.3 & \nodata& 12 12 40 & $-$62 52 & \ \ 5? & ? &  \ \ 5? & 1  \\
298.6$-$0.0 & \nodata& 12 13 41 & $-$62 37 & 12$\times$9 \ & S & \ \ 5? & 2\\
299.2$-$2.9 & \nodata & 12 15 13 & $-$65 30 & 18$\times$11 & S & \ \ \ \ 0.5? & 1\\
%\vspace{-0.2cm}
&&&&&&&\\
299.6$-$0.5 &  \nodata& 12 21 45 & $-$63 09 & 13 & S & \ \ \ \ 1.0? & 1\\
301.4$-$1.0 & \nodata & 12 37 55 & $-$63 49 & 37$\times$23 & S & \ \ \ \ 2.1? & 2 \\
302.3+0.7 & \nodata & 12 45 55 & $-$62 08 & 17 & S & \ \ 5? & 2\\
304.6+0.1 & Kes 17 & 13 05 59 & $-$62 42 & \ 8 & S & 14 \ & 1\\
308.1$-$0.7 & \nodata & 13 37 37 & $-$63 04 & 13 & S & \ \ \ \ 1.2? & 1\\
%\vspace{-0.2cm}
&&&&&&&\\
308.8$-$0.1 & \nodata & 13 42 30 & $-$62 23 & \ 30$\times$20? & \ C? & 15? & 1\\
309.2$-$0.6 & \nodata & 13 46 31 & $-$62 54 & 15$\times$12 & S & \ \ 7? & 1\\
309.8+0.0 & \nodata & 13 50 30 & $-$62 05 & 25$\times$19 & S & 17 \ & 1\\
310.6$-$0.3 & Kes 20B & 13 58 00 & $-$62 09 & \ 8 & S & \ \ 5? & 1\\
310.8$-$0.4 & Kes 20A & 14 00 00 & $-$62 17 & 12 & S & \ \ 6? & 1\\
%\vspace{-0.2cm}
&&&&&&&\\
311.5$-$0.3 & \nodata & 14 05 38 & $-$61 58 & \ 5 & S & \ \ 3? & 2\\
312.4$-$0.4 & \nodata & 14 13 00 & $-$61 44 & 38 & S & 45 \ & 2 \\
315.4$-$2.3 & RCW 86, MSH 14$-$63 & 14 43 00 & $-$62 30 & 42 & S & 49 \ & 2\\
315.4$-$0.3 & \nodata & 14 35 55 & $-$60 36 & 24$\times$13 & ? & \ 8 & 2 \\
315.9$-$0.0 & \nodata & 14 38 25 & $-$60 11 & 25$\times$14 & S & \ \ \ \ 0.8? & 1 \\
%\vspace{-0.2cm}
&&&&&&&\\
&&&&&&&\\
&&&&&&&\\
&&&&&&&\\
&&&&&&&\\
316.3$-$0.0 & (MSH 14$-$57) & 14 41 30 & $-$60 00 & 29$\times$14 & S & 20? & 1 \\
317.3$-$0.2 &  \nodata& 14 49 40 & $-$59 46 & 11 & S & \ \ \ \ 4.7? & 1\\
318.2+0.1 & \nodata & 14 54 50 & $-$59 04 & 40$\times$35 & S & $ \ >3.9$? & 2\\
318.9+0.4 & \nodata & 14 58 30 & $-$58 29 & 30$\times$14 & C & \ \ 4? & 1\\
320.4$-$1.2 & MSH 15$-$52,RCW 89 & 15 14 30 & $-$59 08 & 35 & C & 60? & 2\\

&&&&&&&\\
%\vspace{-0.2cm}
%&&&&&&&\\
320.6$-$1.6 & \nodata & 15 17 50 & $-$59 16 & 60$\times$30 & S & \ ? & 3  \\
321.9$-$1.1 & \nodata & 15 23 45 & $-$58 13 & 28 & S & $ \ >3.4$? & 1\\
321.9$-$0.3 & \nodata & 15 20 40 & $-$57 34 & 31$\times$23 & S & 13 & 3 \\
322.5$-$0.1 & \nodata & 15 23 23 & $-$57 06 & 15 & C & \ \ \ 1.5 & 1\\
323.5+0.1 & \nodata & 15 28 42 & $-$56 21 & 13 & S & \ \ 3? & 2\\
%\vspace{-0.2cm}
&&&&&&&\\
326.3$-$1.8 & MSH 15$-$56 & 15 53 00 & $-$56 10 & 38 & C & 145 &1\\
327.1$-$1.1 &  \nodata& 15 54 25 & $-$55 09 & 18 & C & \ \ 7? & 1\\
327.4+0.4 & Kes 27 & 15 48 20 & $-$53 49 & 21 & S & 30? & 1 \\
327.4+1.0 & \nodata & 15 46 48 & $-$53 20 & 14 & S & \ \ \ \ 1.9? & 1\\
328.4+0.2 & (MSH 15$-$57) & 15 55 30 & $-$53 17 & \ 5 & F & 15 \ & 1\\
&&&&&&&\\
329.7+0.4 & \nodata & 16 01 20 & $-$52 18 & 40$\times$33 & S & $>34$? \ \ & 2 \\
330.2+1.0 & \nodata & 16 01 06 & $-$51 34 & 11 & \ S? & \ \ 5? & 1\\
332.0+0.2 &\nodata  & 16 13 17 & $-$50 53 & 12 & S & \ \ 8? & 1\\
332.4$-$0.4 & RCW 103 & 16 17 33 & $-$51 02 & 10 & S & 28 \ & 2\\
332.4+0.1 & MSH 16$-$51, Kes 32 &16 15 17 & $-$50 42 & 15 & S & 26 \ & 3\\
&&&&&&&\\
335.2+0.1 & \nodata & 16 27 45 & $-$48 47 & 21 & S & 16 \ & 3 \\
336.7+0.5 & \nodata & 16 32 11 & $-$47 19 & 14$\times$10 & S & \ 6 & 1\\
337.0$-$0.1 & (CTB 33) & 16 35 57 & $-$47 36 & \ \ \ \ 1.5 & S & \ \ \ 1.5 & 1\\
337.2$-$0.7 & \nodata & 16 39 28 & $-$47 51 & \ 6 & S & \ \ 2? & 1\\
337.3+1.0 & Kes 40 & 16 32 39 & $-$46 36 & 15$\times$12 & S & 16 \ & 1\\
&&&&&&&\\
337.8$-$0.1 & Kes 41 & 16 39 01 & $-$46 59 & 9$\times$6 & S & 18 \ & 1\\
338.1+0.4 &  \nodata& 16 37 59 & $-$46 24 & \ 15? & S & \ \ 4? & 1\\
338.3$-$0.0 & \nodata & 16 41 00 & $-$46 34 & \ 8 & S & \ \ 7? &2 \\
338.5+0.1 &  \nodata& 16 41 09 & $-$46 19 & \ 9 & ? & 12? &3 \\
340.4+0.4 & \nodata & 16 46 31 & $-$44 39 & 10$\times$7 \ & S & \ 5 & 1\\
&&&&&&&\\
340.6+0.3 & \nodata & 16 47 41 & $-$44 34 & \ 6 & S & \ \ 5? & 1\\
341.2+0.9 & \nodata & 16 47 35 & $-$43 47 & 16$\times$22 & C & \ \ \ \ 1.5? &1 \\
341.9$-$0.3 &  \nodata& 16 55 01 & $-$44 01 & \ 7 & S & \ \ \ 2.5 & 1 \\
342.0$-$0.2 & \nodata & 16 54 50 & $-$43 53 & 12$\times$9 \ & S & \ \ \ \ 3.5? & 1 \\
342.1+0.9 & \nodata& 16 50 43 & $-$43 04 & 10$\times$9 \ & S & \ \ \ \ 0.5? & 1\\
&&&&&&&\\
&&&&&&&\\
&&&&&&&\\
&&&&&&&\\
&&&&&&&\\
343.0$-$6.0 & \nodata& 17 25 00 & $-$46 30 & 250 \ & S & \ ? & 1\\
343.1$-$2.3 &  \nodata& 17 08 00 & $-$44 16 & \ 32? & \ C? & \ \ 8? &3 \\
343.1$-$0.7 & \nodata & 17 00 25 & $-$43 14 & 27$\times$21 & S & \ \ \ 7.8 & 3 \\
344.7$-$0.1 & \nodata & 17 03 51 & $-$41 42 & 10 & \ C? & \ \ \ \ 2.5? & 1\\
345.7$-$0.2 & \nodata & 17 07 20 & $-$40 53 & \ 6 & S & \ \ \ \ 0.6? & 1\\
&&&&&&&\\
346.6$-$0.2 & \nodata & 17 10 19 & $-$40 11 & \ 8 & S & \ \ 8? & 1\\
347.3$-$0.5 & \nodata & 17 13 50 & $-$39 45 & 65$\times$55 & \ S? & \ ? & 3\\
348.5$-$0.0 &  \nodata& 17 15 26 & $-$38 28 & \ 10? & \ S? & 10? & 1\\
348.5+0.1 & CTB 37A & 17 14 06 & $-$38 32 & 15 & S & 72 \ & 1\\
348.7+0.3 & CTB 37B & 17 13 55 & $-$38 11 & \ 17? & S & 26 \ & 1\\
%&&&&&&&\\
&&&&&&&\\
349.2$-$0.1 & \nodata & 17 17 15 & $-$38 04 & 9$\times$6 & S & \ \ \ \ 1.4? & 1\\
349.7+0.2 & \nodata & 17 17 59 & $-$37 26 & 2.5$\times$2 \ \ & S & 20 \ & 2\\
350.0$-$2.0 & \nodata & 17 27 50 & $-$38 32 & 45 & S & 26 \ & 3\\
351.2+0.1 & \nodata & 17 22 27 & $-$36 11 & \ 7 & \ C? & \ \ 5? & 2\\
351.7+0.8 & \nodata & 17 21 00 & $-$35 27 & 18$\times$14 & S & 10? & 1\\
&&&&&&&\\
351.9$-$0.9 & \nodata & 17 28 52 & $-$36 16 & 12$\times$9 \ & S & \ \ \ \ 1.8? & 1\\
352.7$-$0.1 & \nodata & 17 27 40 & $-$35 07 & 8$\times$6 & S & \ 4 & 2\\
353.9$-$2.0 & \nodata & 17 38 55 & $-$35 11 & 13 & S & \ \ 1? & 1\\
354.1+0.1 & \nodata & 17 30 28 & $-$33 46 & \ 15$\times$3? \ & \ C? & \ ? & 1\\
354.8$-$0.8 & \nodata & 17 36 00 & $-$33 42 & 19 & S & \ \ \ \ 2.8? & 1\\
&&&&&&&\\
355.6$-$0.0 & \nodata & 17 35 16 & $-$32 38 & 8$\times$6 & S & \ \ 3? & 1\\
355.9$-$2.5 & \nodata & 17 45 53 & $-$33 43 & 13 & S & \ 8 & 1\\
356.2+4.5 & \nodata & 17 19 00 & $-$29 40 & 25 & S & \ 4 & 1\\
356.3$-$0.3 & \nodata & 17 37 56 & $-$32 16 & 11$\times$7 \ & S & \ \ 3? & 2\\
356.3$-$1.5 & \nodata & 17 42 35 & $-$32 52 & 20$\times$15 & S & \ \ 3? &1\\
&&&&&&&\\
357.7$-$0.1 & \nodata & 17 40 29 & $-$30 58 & 8$\times$3 & ? & 37\ &1\\
357.7$+$0.3 & \nodata & 17 38 35 & $-$30 44 & 24 & S & 10\ &2\\
%\hline
%\end{tabular}
%\end{center} \end{table}
\enddata
\end{deluxetable}
\clearpage

\begin{deluxetable}{ccrc}
\tabletypesize{\scriptsize}
\tablecaption{Excess Emissions at Rank 2 SNRs\label{tbl$$-$$3}}
\tablewidth{0pt}
\tablehead{
\colhead{Galactic} & \colhead{$v_{\rm min}$, $v_{\rm max}$}   &
\colhead{$\int \Delta \bar{T}_{A} \,dv$}& \colhead{}\\
\colhead{Coordinates} & \colhead{(\kms)} & \colhead{(K~\kms)}& \colhead{Note}
}
\startdata

286.5$-$1.2&$-$52.5, $-$28.6 & 122.3$\pm$10.9 & 1\\
294.1$-$0.0& $+$60.4, $+$71.1 & 70.9$\pm$2.7 & 1\\
        & +29.0, +40.6& 130.0$\pm$12.6& 3\\
        &\ $-$3.9, +10.9& 187.5$\pm$13.6& 3\\
        & $-$29.5, $-$18.7& 69.2$\pm$8.3& 3\\
296.1$-$0.5& +48.0, +62.9& 174.8$\pm$2.0 & 2\\
        & +10.1, +23.3& 290.2$\pm$25.1& 2\\
296.8$-$0.3&+35.7, +46.4& 286.9$\pm$30.7& 2\\
298.6$-$0.0&+15.9, +28.3& 314.2$\pm$37.1 & 3\\
301.4$-$1.0&+11.8, +23.3& 250.4$\pm$25.9&1\\
302.3+0.7&+44.7, +56.3& 108.2$\pm$10.0& 2\\
311.5$-$0.3& $-$15.4, $-$1.4 \ & 232.9$\pm$25.0& 3\\
312.4$-$0.4&+12.6, +34.0& 532.8$\pm$19.8 & 3\\
315.4$-$2.3&+66.2, +78.5& 116.1$\pm$5.0 &3(?)\\
315.4$-$0.3& \ $-$0.6, +17.5& 657.3$\pm$46.8&3\\
318.2+0.1& \ $-$2.2, +21.7& 314.3$\pm$14.5& 2\\
        &$-$65.7, $-$49.2& 248.3$\pm$14.0 & 3\\
320.4$-$1.2& \ +5.2, +18.3& 217.7$\pm$11.1&3\\
323.5+0.1&$-$110.3, $-$93.8 \ & 94.3$\pm$7.8 &2\\
329.7+0.4&+66.2, +78.5& 34.7$\pm$2.4 &1(?)\\
        & \ +5.2, +15.9& 75.4$\pm$6.9&3\\
        &$-$54.2, $-$38.5& 172.2$\pm$11.1 &3\\
        &$-$111.1, $-$76.5 \ &659.5$\pm$26.4 &3\\
332.4$-$0.4&$-$40.2, $-$27.0&306.6$\pm$28.1 &3\\
%335.2+0.1&$-$14.6, +2.7 \ &277.0$\pm$27.4 & 1(?), 4\\
338.3$-$0.0&$-$21.2, $-$7.2 \ &355.8$\pm$44.3 &3\\
349.7+0.2&$-$115.2, $-$97.9 \ &489.8$\pm$60.6 &3\\
351.2+0.1&$-$210.0, $-$196.8&83.3$\pm$8.6&3\\
352.7$-$0.1&$-$61.6, $-$46.8&413.5$\pm$50.6 &3\\
356.3$-$0.3&$-$244.6, $-$233.9&18.2$\pm$3.4&3\\
357.7$+$0.3&104.7, 127.8&166.2$\pm$25.1& 3\\
     &56.1, 74.2&40.3$\pm$8.9& 3\\
     &$-$17.3, $-$5.8&153.6$\pm$23.3& 3\\
     &$-$86.6, $-$74.2&73.3$\pm$16.3& 3\\
&$-$244.0, $-$147.6&267.2$\pm$15.5& 3\\
\enddata

\tablecomments{(1) Excess emission confined to a small area centered at SNR.
(2) Excess emission confined to a small area near SNR boundary.
(3) Excess emission extended over a large area surrounding SNR.
%(4) Also belongs to rank 3 SNRs.
}

\end{deluxetable}

\begin{deluxetable}{ccrc}
\tabletypesize{\scriptsize}
\tablecaption{High-Velocity Excess Emission at Rank 3 SNRs\label{tbl$-$2}}
\tablewidth{0pt}
\tablehead{
\colhead{Galactic} & \colhead{$v_{\rm min}$,$v_{\rm max}$}   &
\colhead{$\int \Delta \bar{T}_{A} \,dv$}& \colhead{}\\
\colhead{Coordinates} & \colhead{(\kms)} & \colhead{(K~\kms)}& \colhead{Note}
}
\startdata

%\newcolumntype{,}{D{,}{,}{2}}
%\newcolumntype{+}{D{$\pm$}{$\pm$}{2}}
%\begin{table}[!t]
%\footnotesize
%%\tiny
%\begin{center}
%\caption{The Catalog of the Forbidden-Velocity Wings (1)}
%\vspace{0.3cm}
%\begin{tabular}{c c c c}
%%\begin{tabular}{c , + c}
%\hline\hline
%Galactic&$v_{\rm min}$, \ $v_{\rm max}$&
%$\int \Delta \bar{T}_{A} \,dv$&		 \\
%Coordinates&	kms&K kms& Notes\\
%\hline

272.2$-$3.2& $-$93.8, $-$69.9 & 4.6$\pm$0.6&1\\
320.6$-$1.6& $+$100.8, $+$120.6 & 10.4$\pm$0.7 &3\\
321.9$-$0.3& $-$162.2, $-$108.6 & 8.7$\pm$0.5&1 \\
332.4+0.1& $-$140.7, $-$128.4 & 9.6$\pm$2.6&3 \\
335.2+0.1& $-$154.8, $-$138.3& 3.0$\pm$1.1&3 \\
338.5+0.1& $-$183.6, $-$158.9& 1.7$\pm$0.9&2 \\
343.1$-$2.3& $-$151.5, $-$118.5& 34.0$\pm$2.1&3 \\
343.1$-$0.7& $-$167.1, $-$144.9& 5.6$\pm$1.7&3\\
347.3$-$0.5& $+$67.8, $+$81.0& 9.1$\pm$0.2&1 \\
350.0$-$2.0& $-$184.5, $-$160.5 & 2.4$\pm$0.1& 2 \\
\enddata

%%%% 2 SNRs in L350 is divided with 1.4 .
%%%%
%347.3$-$0.5& $+$67.8, $+$81.0& 12.7$\pm$0.3&3 \\
%350.0$-$2.0& $-$184.5, $-$160.5 & 3.4$\pm$0.2& 1 \\

%\hline

%\tablecomments{Velocity ranges, at which regions of SNRs having
%excessive emissions comparing with surroundings,
%are selected by eyes based on the channel map}
\tablecomments{(1) Excess emission confined to SNR.
(2) Excess emission scattered over an area surrounding SNR. Not enough 
sensitivity. 
(3) Excess emission in confusing area. }
%\end{tabular}
%\end{center}
%\end{table}

\end{deluxetable}

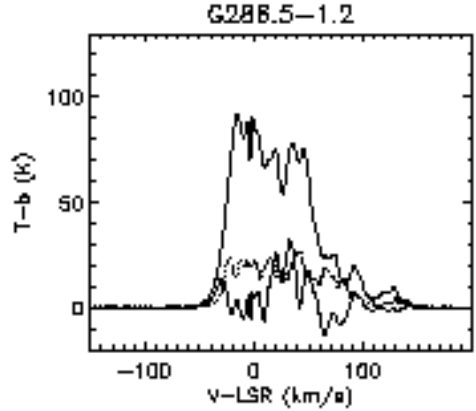
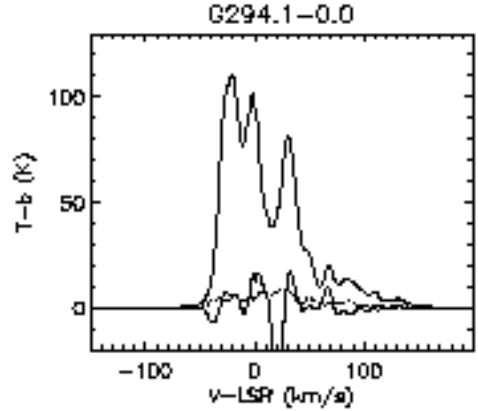
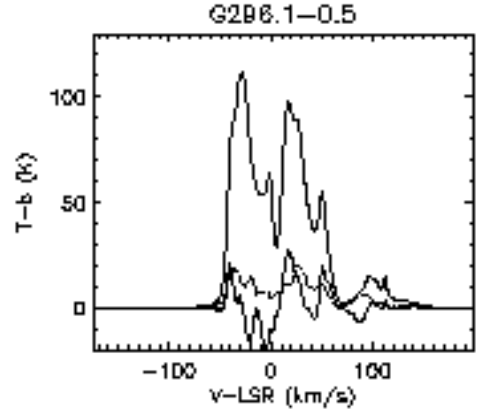
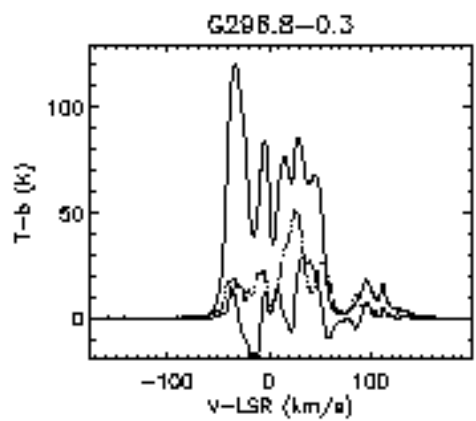
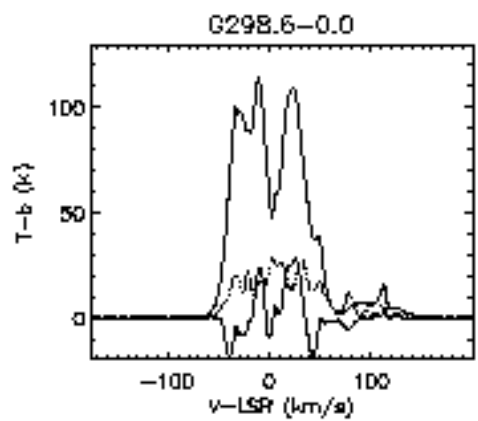
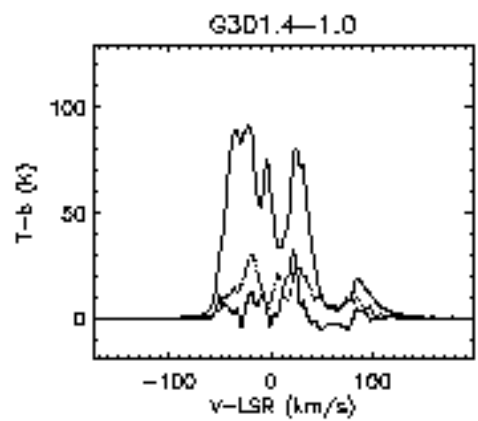
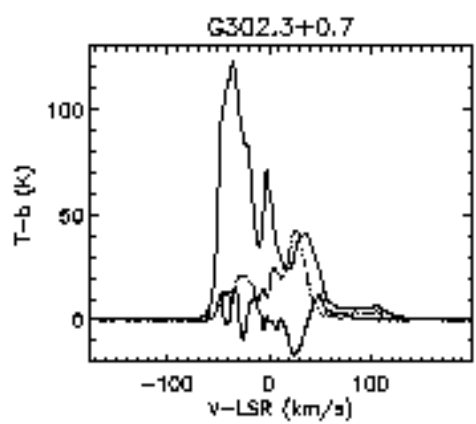
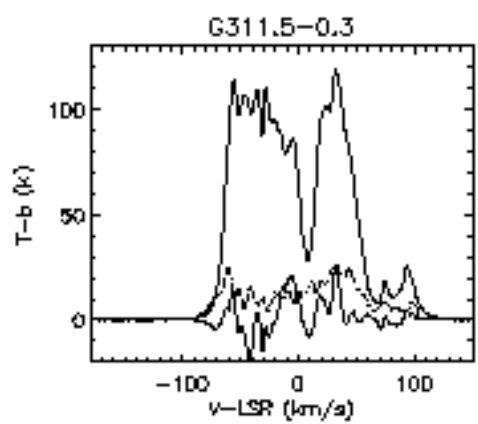
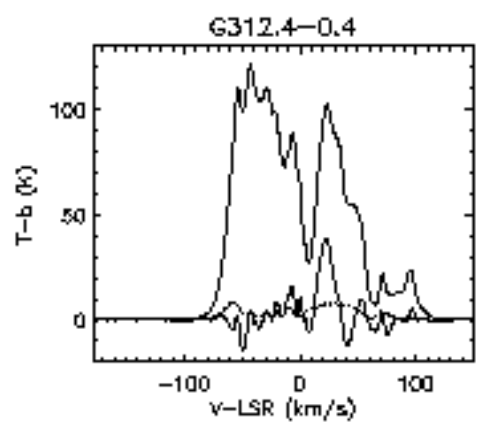
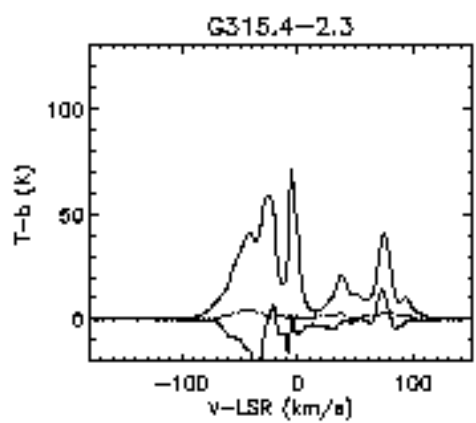
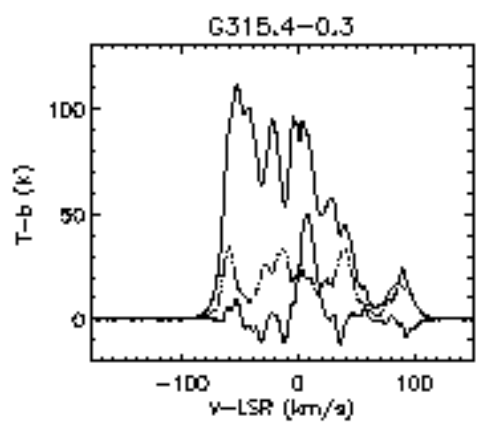
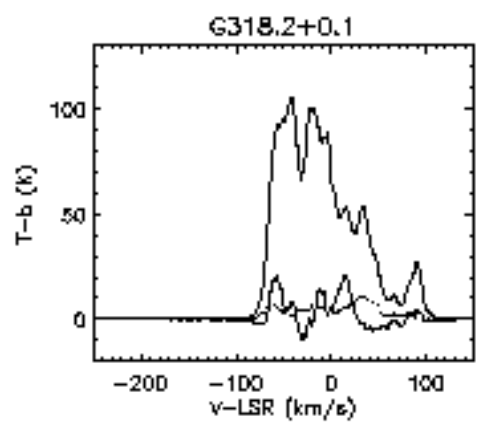
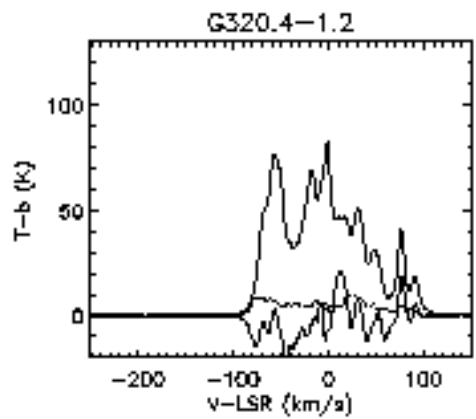
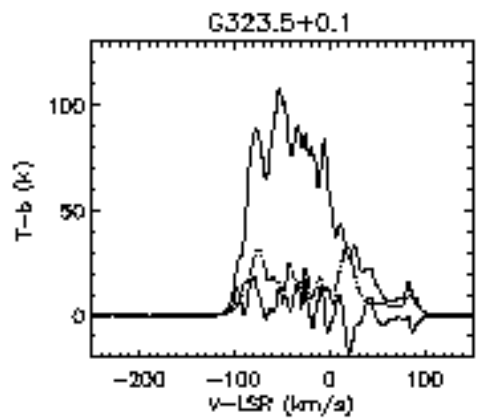
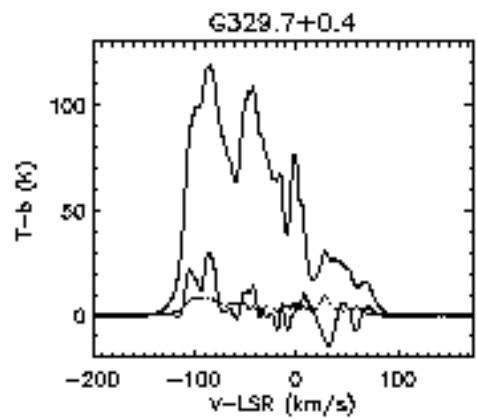
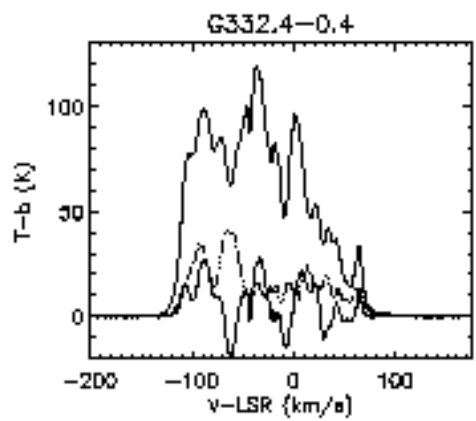
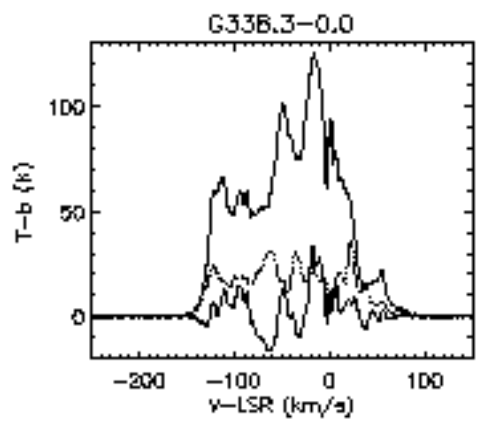
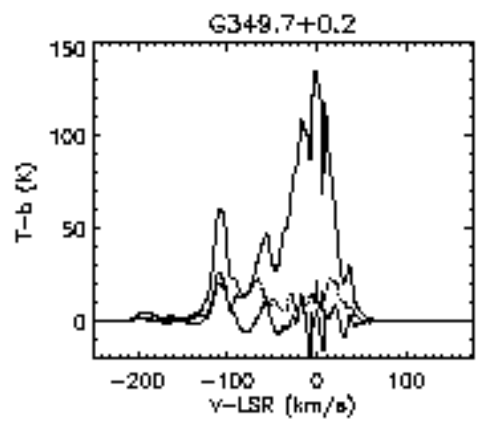
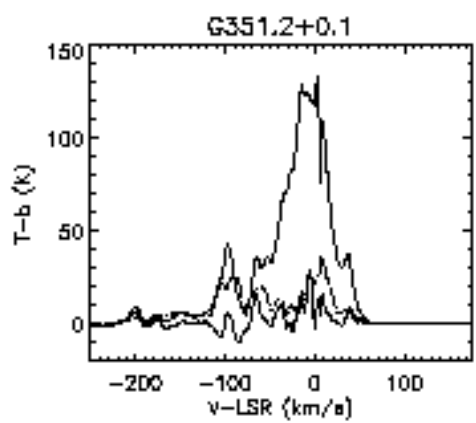
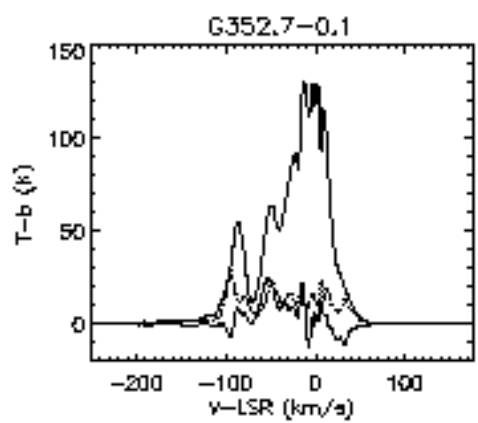
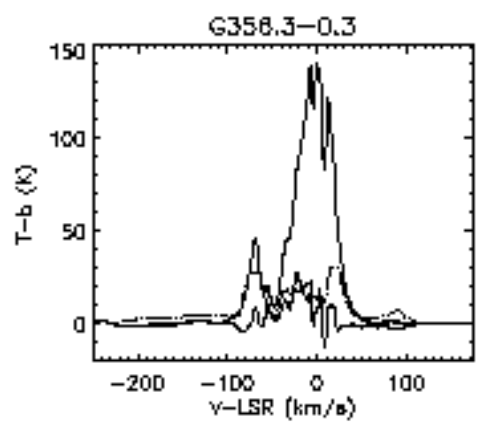
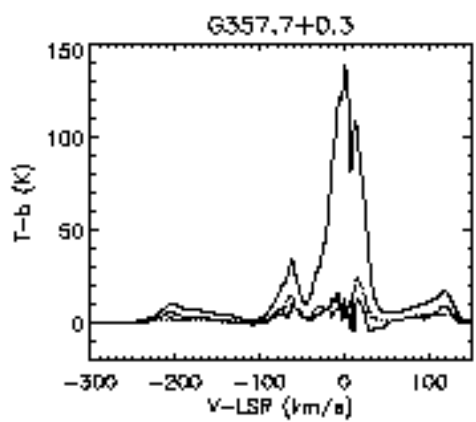
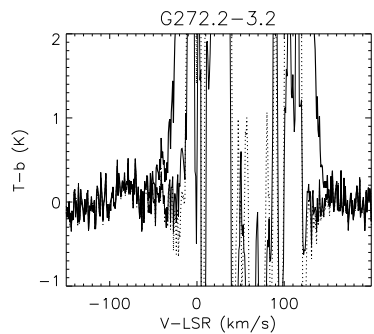
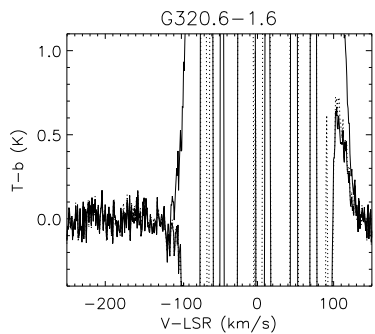
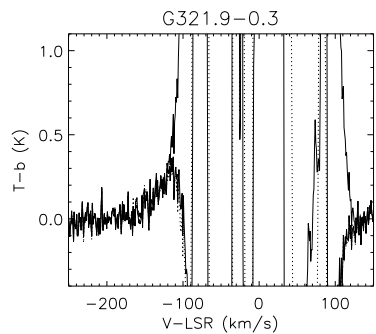
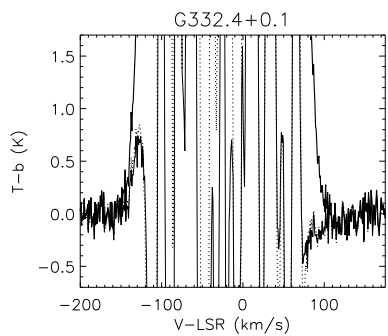
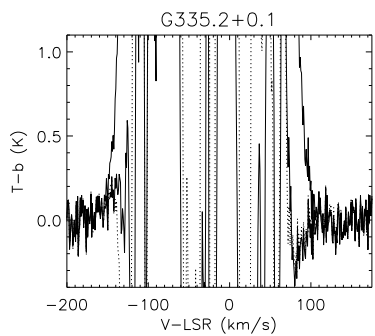
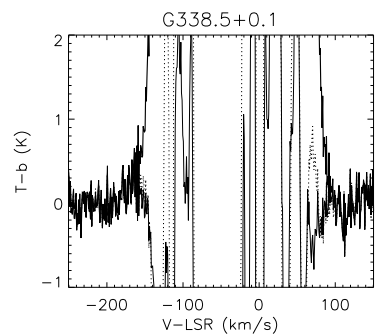
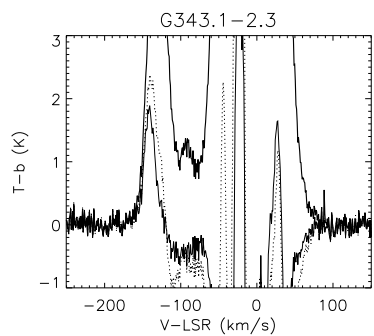
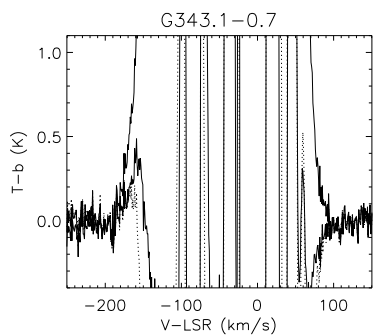
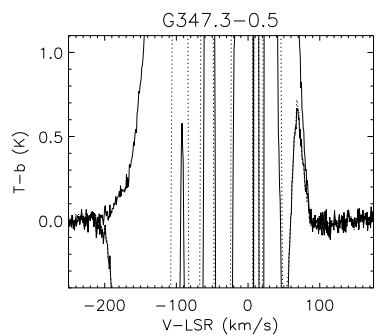
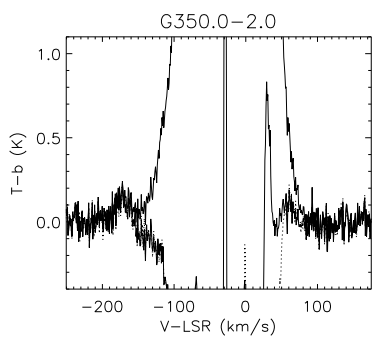
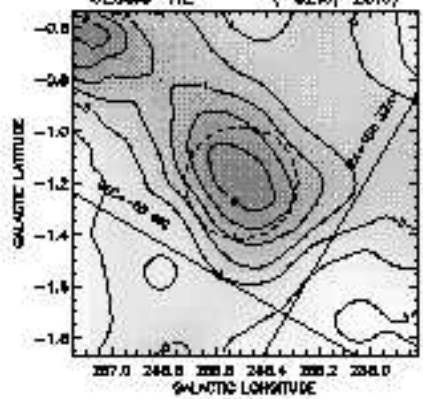
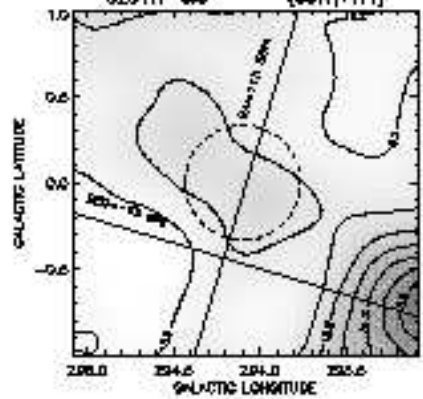
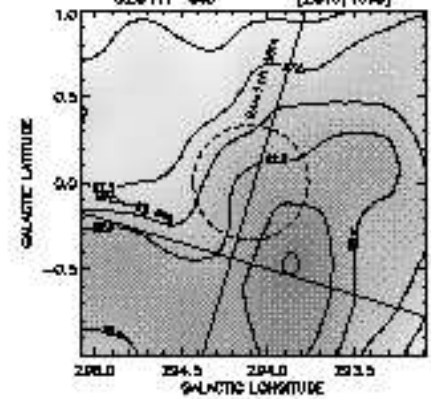
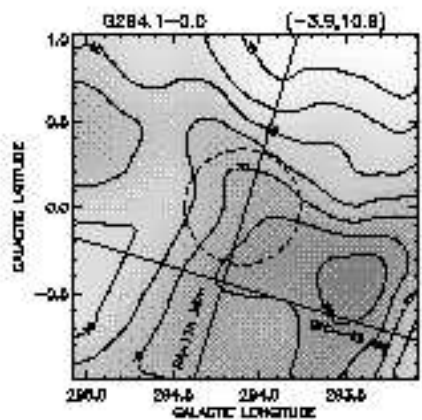
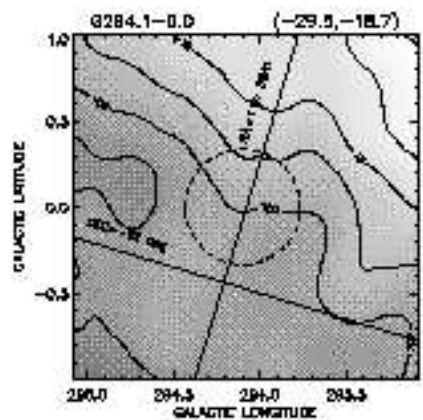
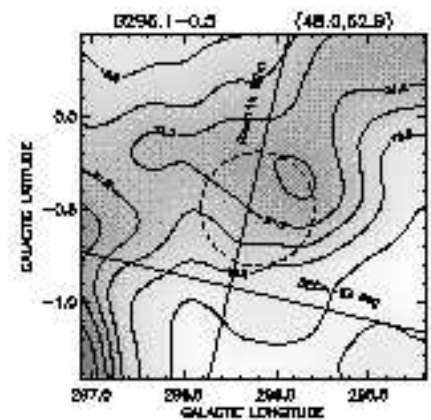
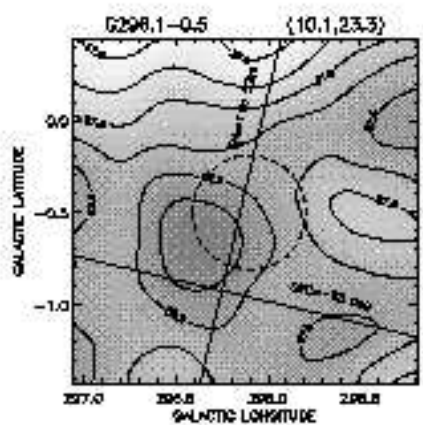
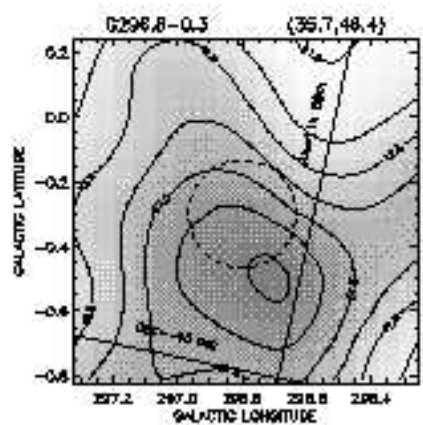
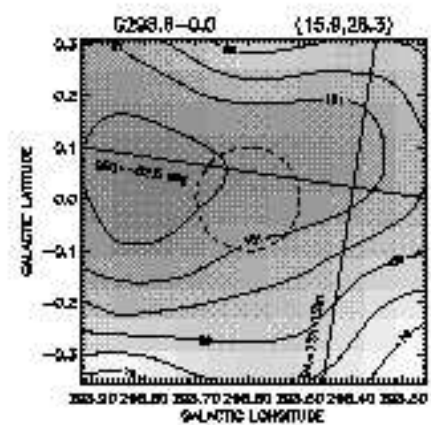
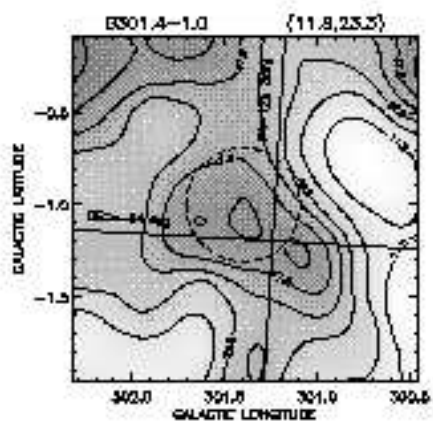
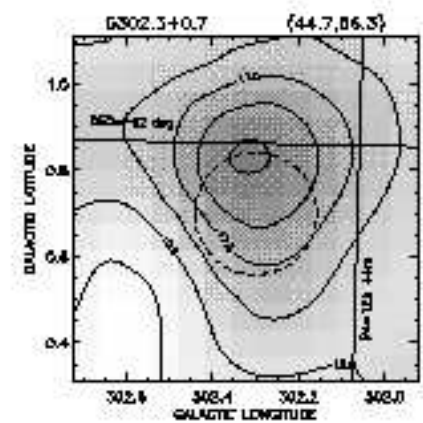
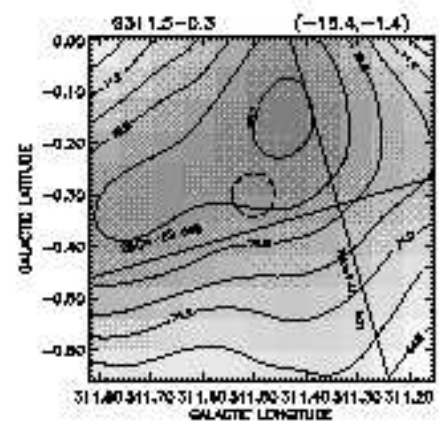
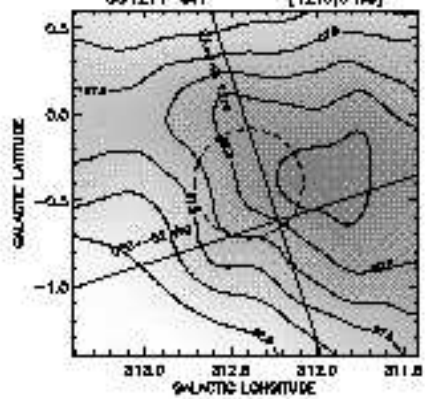
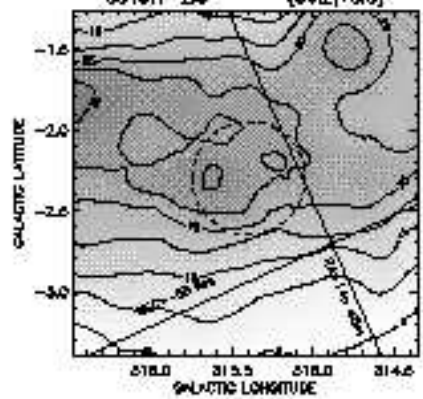
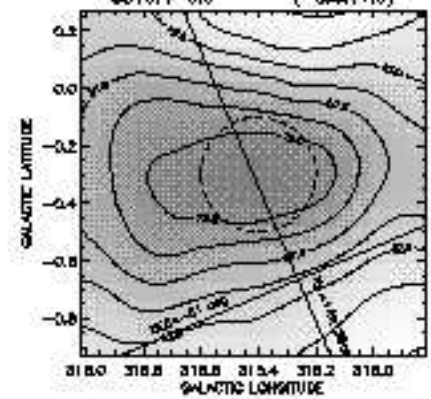
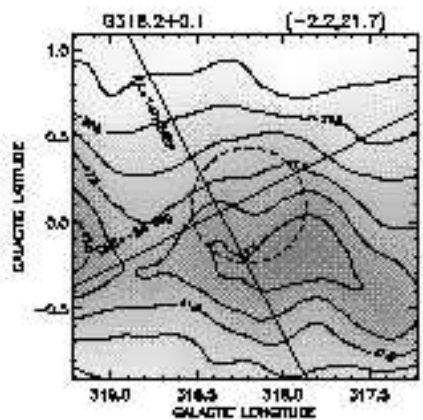
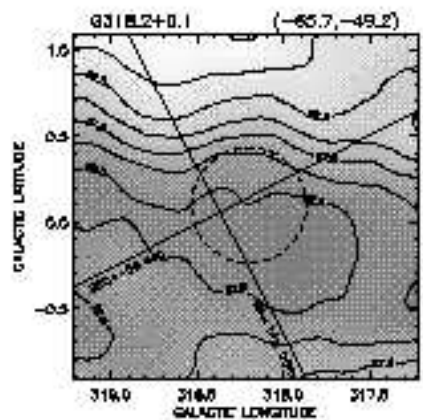
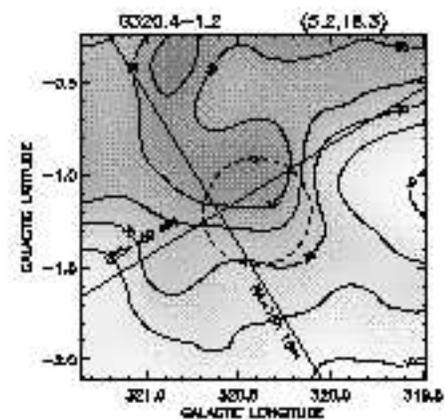
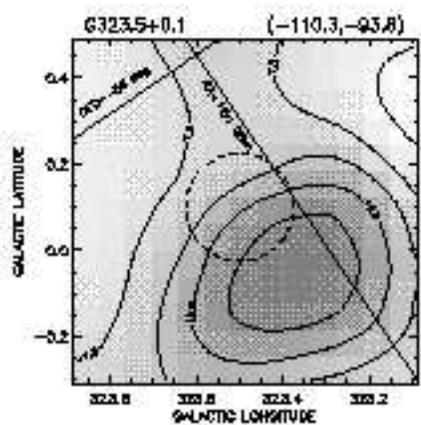
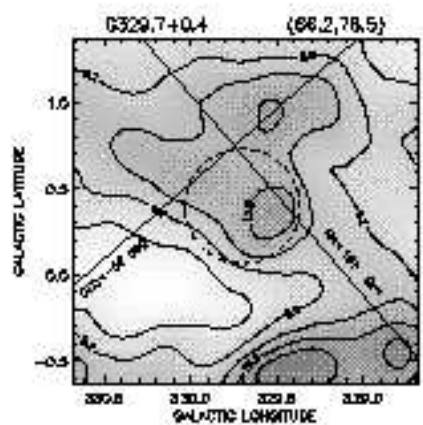
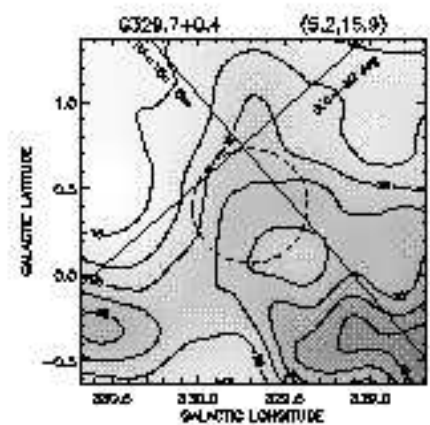
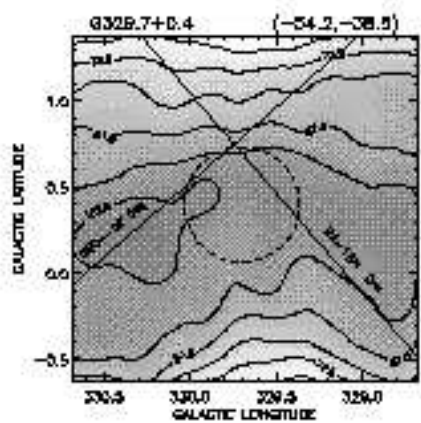
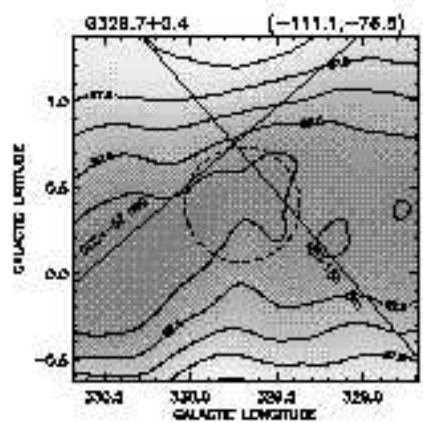
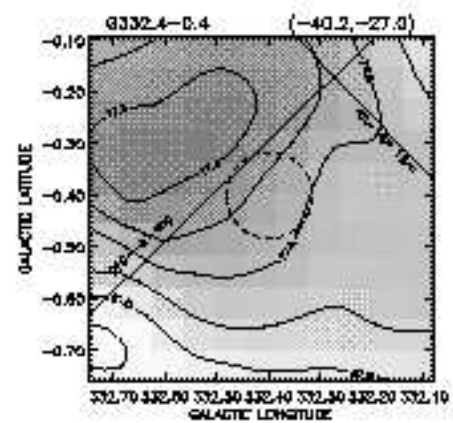
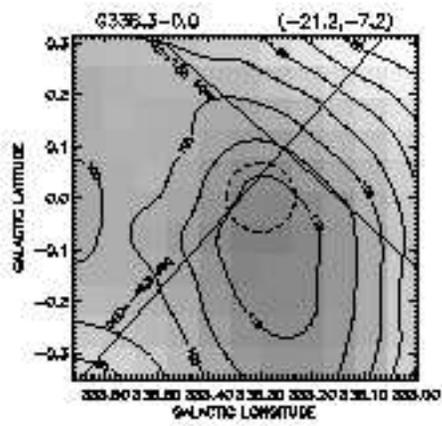
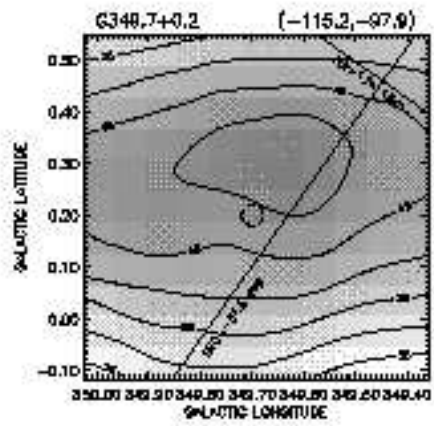
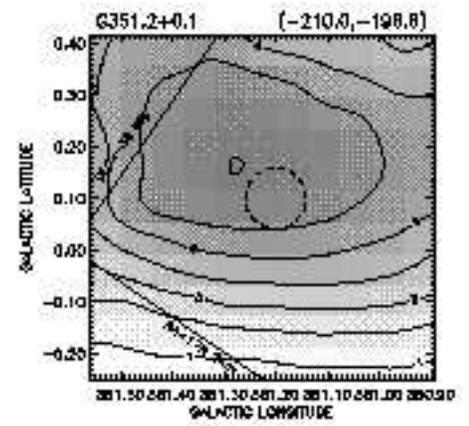
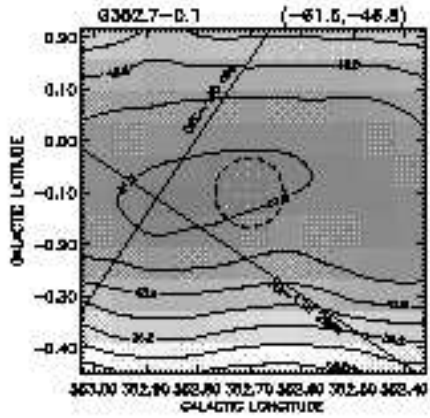
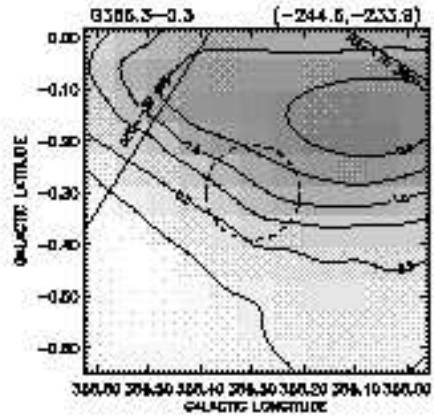
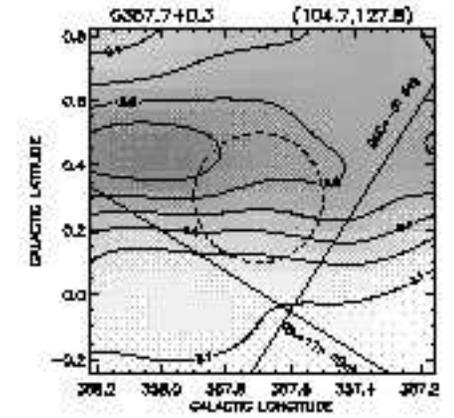
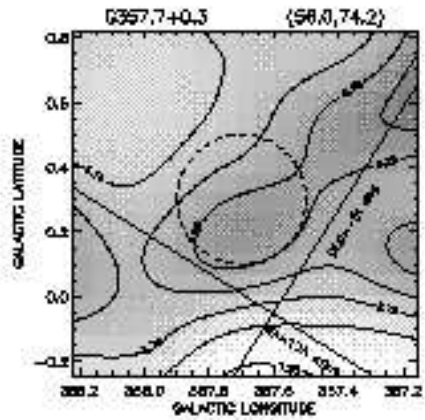
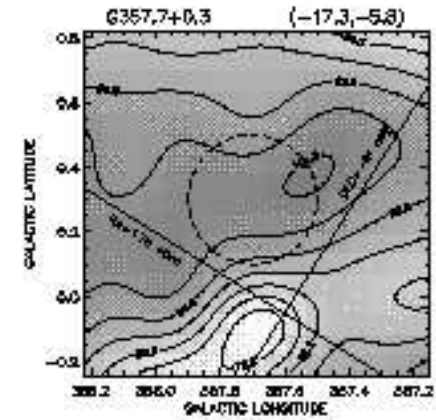
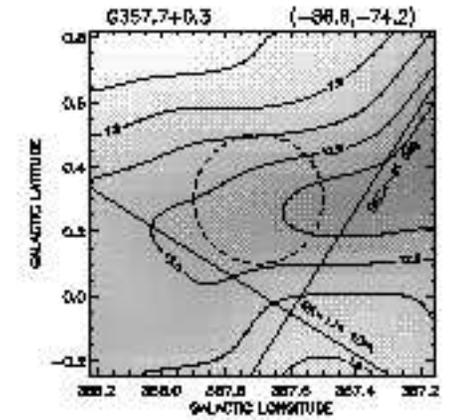
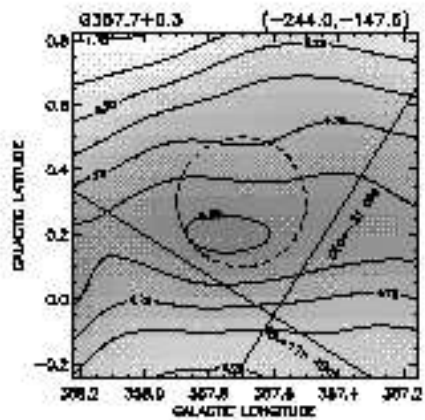
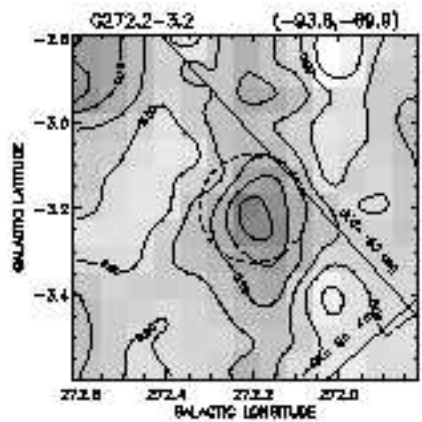
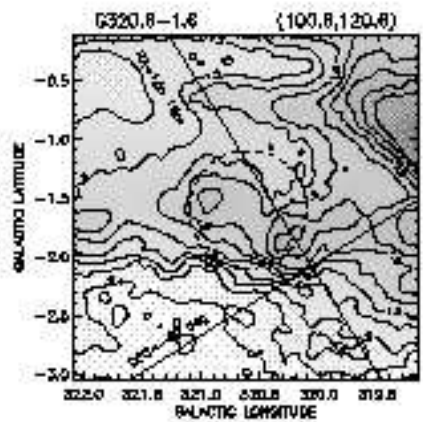
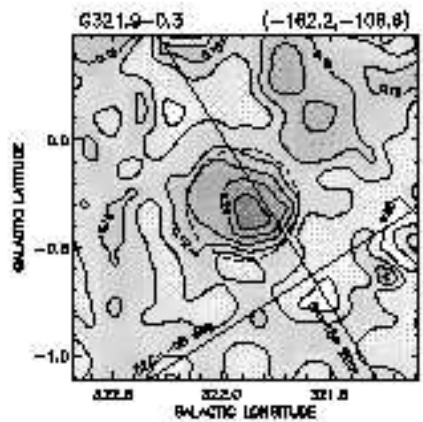
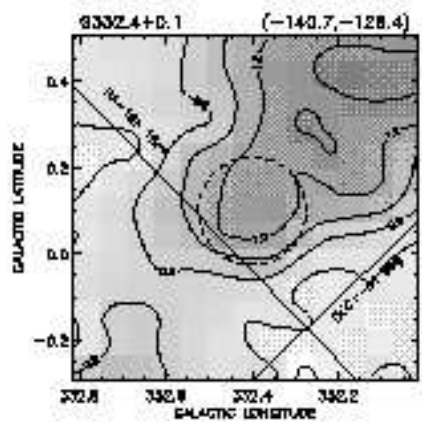
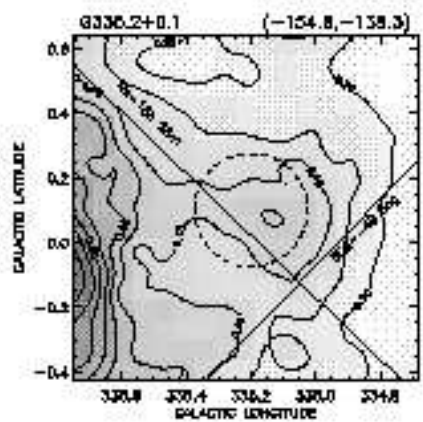
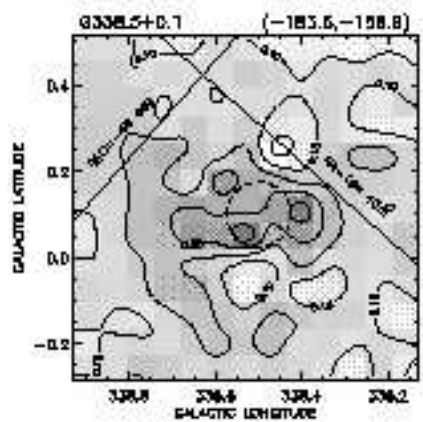
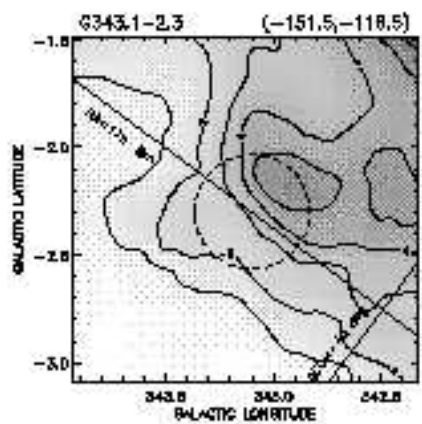
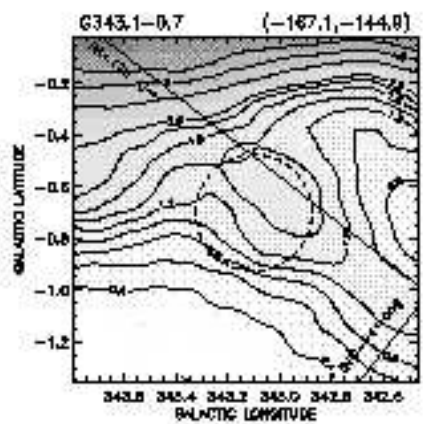
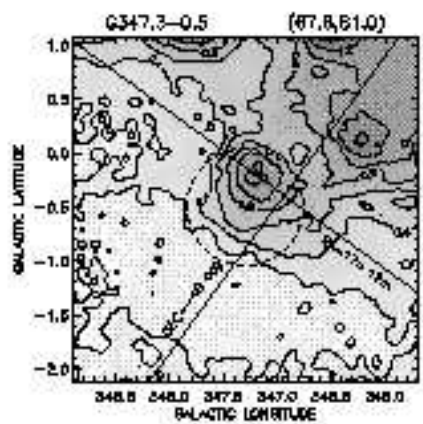
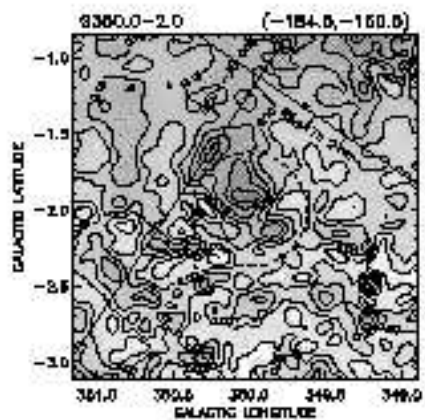
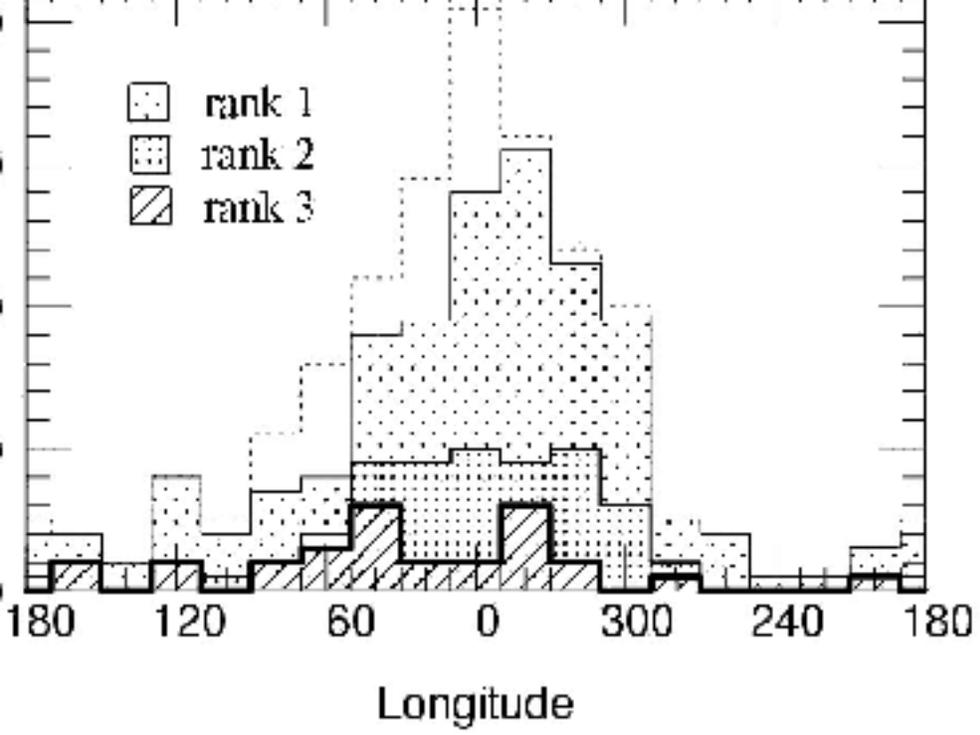
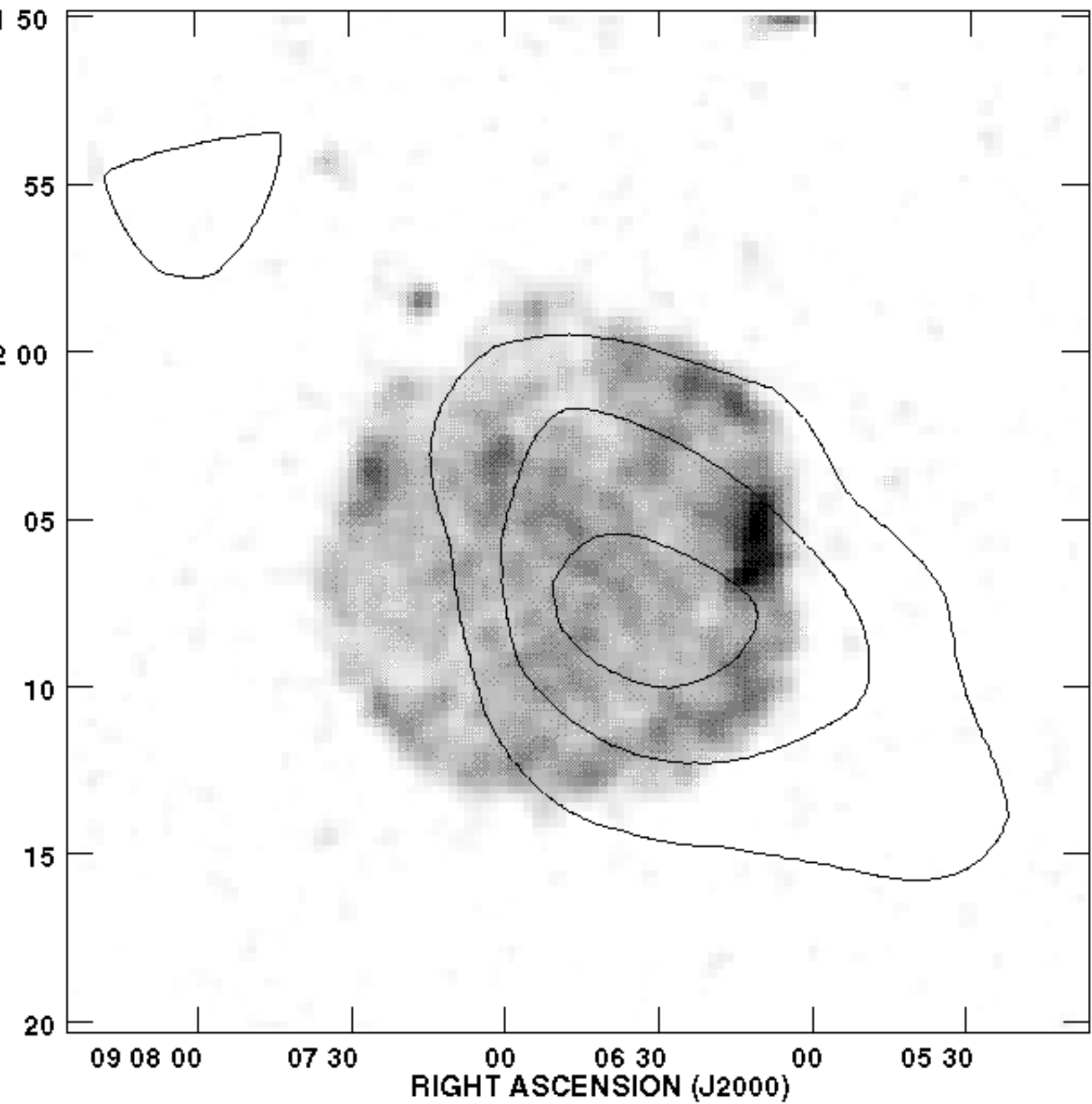
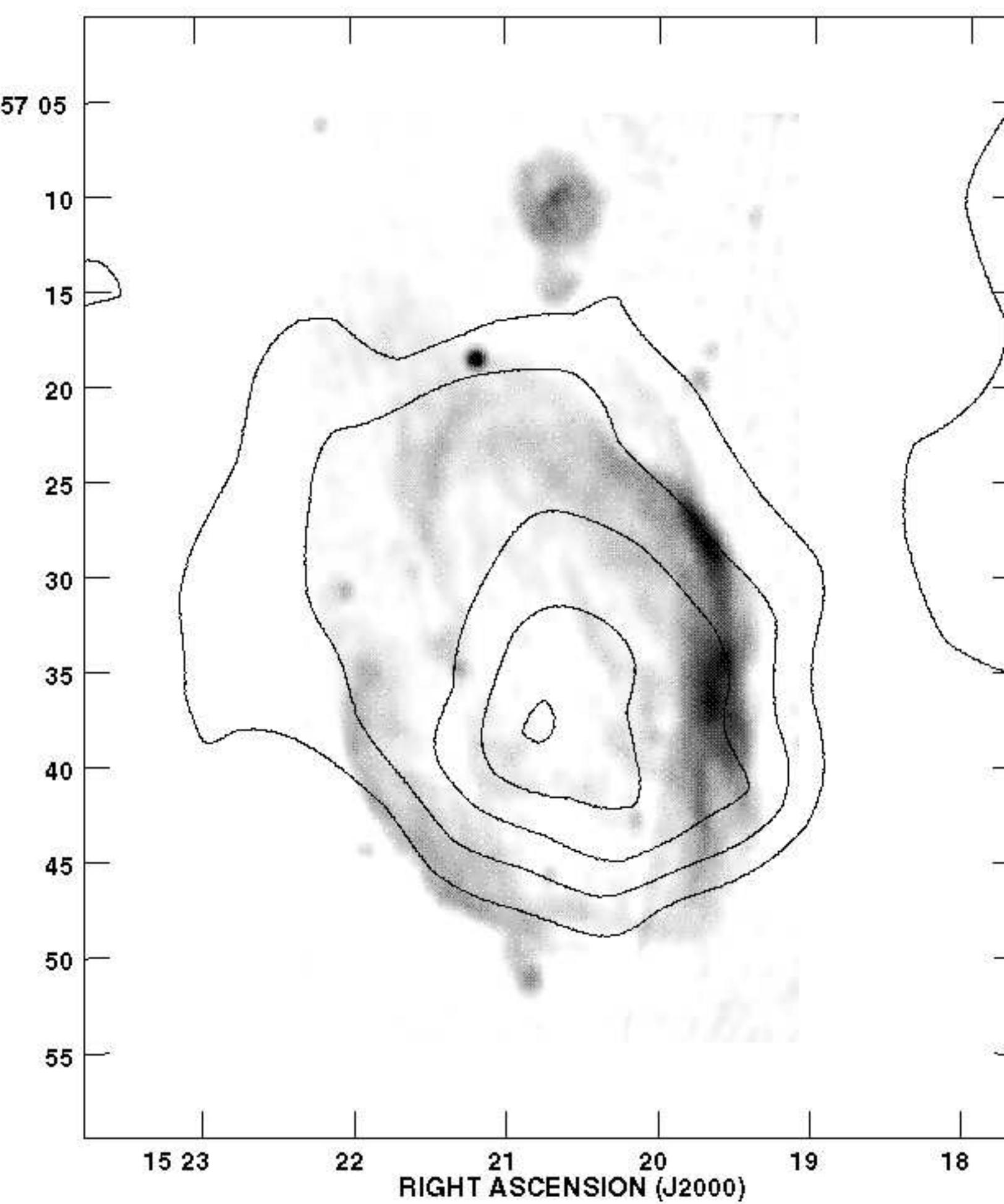
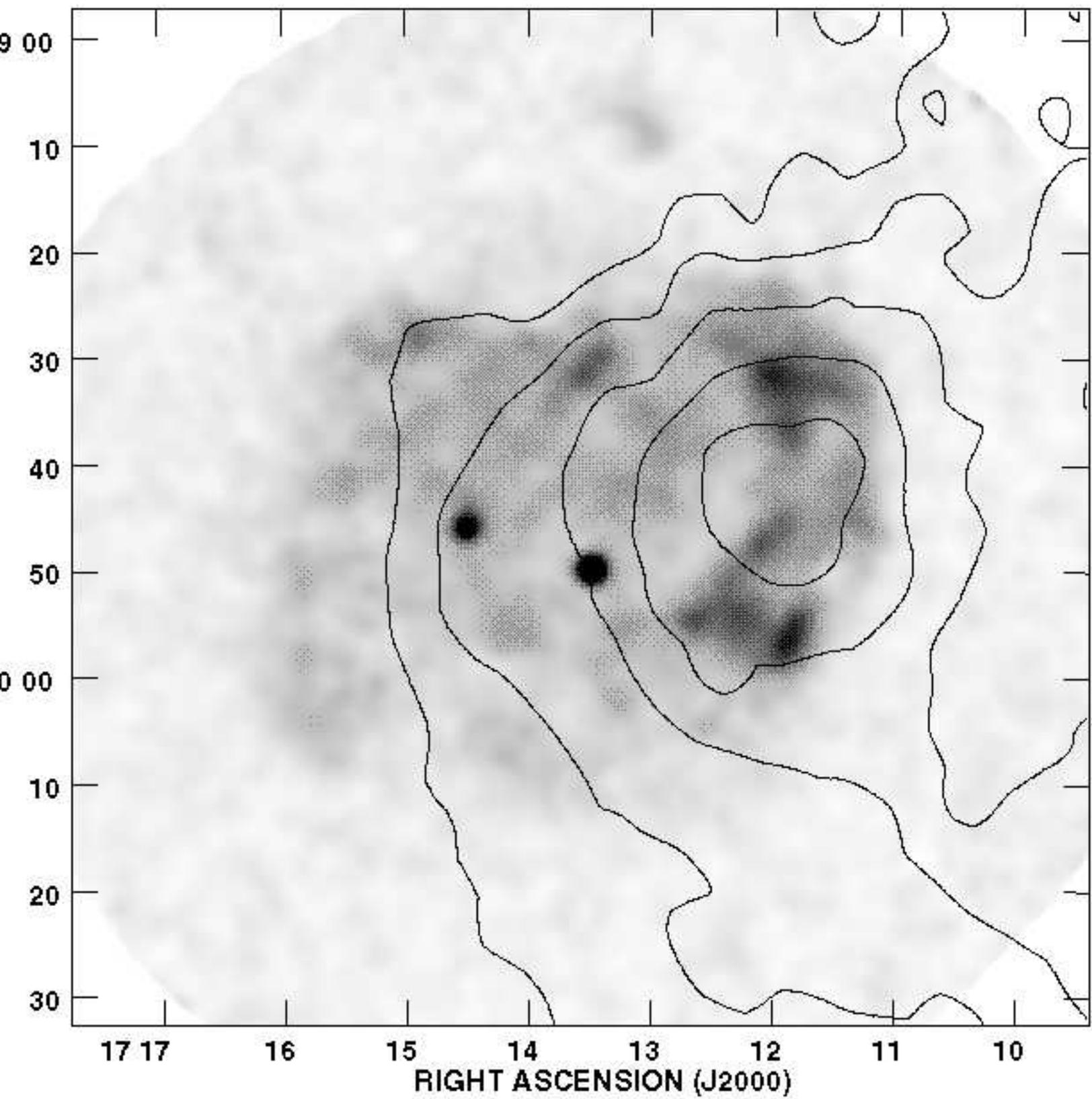
\end{document}